\documentclass[usegraphicx,usenatbib]{mn2e}

\usepackage{amsmath}
\usepackage{amssymb}
\usepackage{times}
\usepackage{subfigure}
\usepackage{epsfig}
\usepackage{txfonts}

\renewcommand{\descriptionlabel}[1]%
 {\hspace{\labelsep}\textbf{#1}}

\begin{document}

\voffset-.6in

\title[Substructure revealed by RR Lyraes in SDSS Stripe 82] {
  Substructure revealed by RR Lyraes in SDSS Stripe 82}

\author[L.L. Watkins et al.]
  {L.L.~Watkins$^1$, N.W. Evans$^1$, V. Belokurov$^1$, M.C. Smith$^1$,
    P.C. Hewett$^1$, D.M. Bramich$^2$, \newauthor G.F. Gilmore$^1$,
    M.J. Irwin$^1$, S. Vidrih$^{1,3,4}$, \L. Wyrzykowski$^1$, 
    D.B. Zucker$^{1,5,6}$.
 \medskip
 \\$^1$Institute of Astronomy, University of Cambridge, Madingley Road, Cambridge, CB3 0HA, UK
 \\$^2$Isaac Newton Group of Telescopes, Apartado de Correos 321,
 E-38700 Santa Cruz de la Palma, Canary Islands, Spain
 \\$^3$Astronomisches Rechen-Institut, Zentrum f\"ur Astronomie der 
 Universit\"at Heidelberg, M\"onchhofstrasse 12-14, 69120 Heidelberg, Germany
 \\$^4$Faculty of Mathematics and Physics, University of Ljubljana, Ljubljana, Slovenia
 \\$^5$Department of Physics, Macquarie University, North Ryde, NSW 2109, Australia
 \\$^6$Anglo-Australian Observatory, P.O. Box 296, Epping, NSW 1710, Australia}

\date{Accepted 2005 August ???. Received 2005 August ???; Submitted 2005 November 4th}

\pagerange{\pageref{firstpage}--\pageref{lastpage}} \pubyear{2004}

\maketitle

\label{firstpage}

\begin{abstract}
  We present an analysis of the substructure revealed by RR Lyraes in
  Sloan Digital Sky Survey (SDSS) Stripe 82, which covers 2\fdg5 in
  declination on the celestial equator over the right ascension range
  $\alpha =$20.7$^{\mbox{\small h}}$ to 3.3$^{\mbox{\small h}}$. We
  use the new public archive of light-motion curves in Stripe 82,
  published by Bramich et al. in 2008, to identify a set of
  high-quality RR Lyrae candidates.  Period estimates are determined
  to high accuracy using a string-length method. A subset of 178 RR
  Lyraes with spectrally derived metallicities are employed to derive
  metallicity-period-amplitude relations, which are then used,
  together with archive magnitude data and lightcurve Fourier
  decomposition, to estimate metallicities and hence distances for the
  entire sample.  The RR Lyraes lie 5--115\,kpc from the Galactic
  center, with distance estimates accurate to $\sim 8$ per cent. The
  RR Lyraes are further divided into subsets of 316 RRab types and 91
  RRc types based on their period, colour and metallicity.

  We fit a smooth density law to the distribution as a simple representation 
  of the data. For Galactocentric radii 5--25\,kpc the number density of RR 
  Lyraes falls as $r^{-2.4}$, but beyond 25\,kpc, the number density falls 
  much more steeply, as $r^{-4.5}$.  However, we stress that in practice the 
  density distribution is not smooth, but dominated by clumps and 
  substructure.  Samples of 55 and 237 RR Lyraes associated with the 
  Sagittarius Stream and the Hercules-Aquila Cloud respectively are 
  identified. Hence, $\sim$70 per cent of the RR Lyraes in Stripe 82 belong to 
  known substructure, and the sharp break in the density law reflects the fact 
  that the dominant substructure in Stripe 82 -- the Hercules-Aquila Cloud and 
  the Sagittarius Stream -- lie within 40\,kpc. In fact, almost 60 per cent of 
  all the RR Lyraes in Stripe 82 are associated with the Hercules-Aquila Cloud 
  alone, which emphasises the Cloud's pre-eminence. Additionally, evidence of 
  a new and distant substructure -- the {\it Pisces Overdensity} -- is found, 
  consisting of $28$ faint RR Lyraes centered on Galactic coordinates $(\ell 
  \approx 80^\circ, b \approx -55^\circ$), with distances of $\sim$80\,kpc. 
  The total stellar mass in the Pisces Overdensity is $\sim$$10^4$\,M$_\odot$ 
  and its metallicity is [Fe/H] $\sim -1.5$.
\end{abstract}


\section{Introduction}
\label{sect:intro}

\begin{figure*}
\epsfig{file=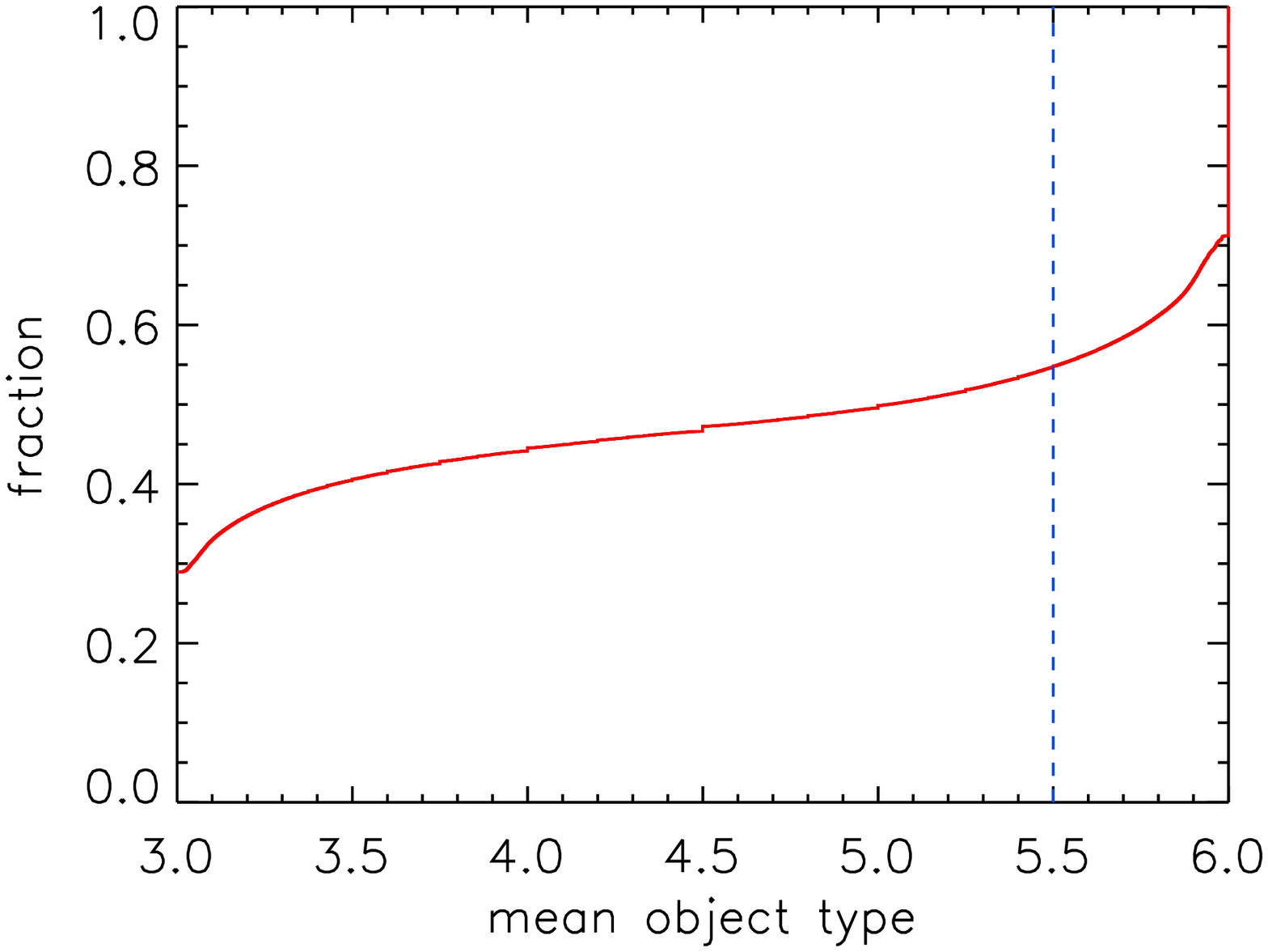,angle=0.0,width=0.4\linewidth}
\epsfig{file=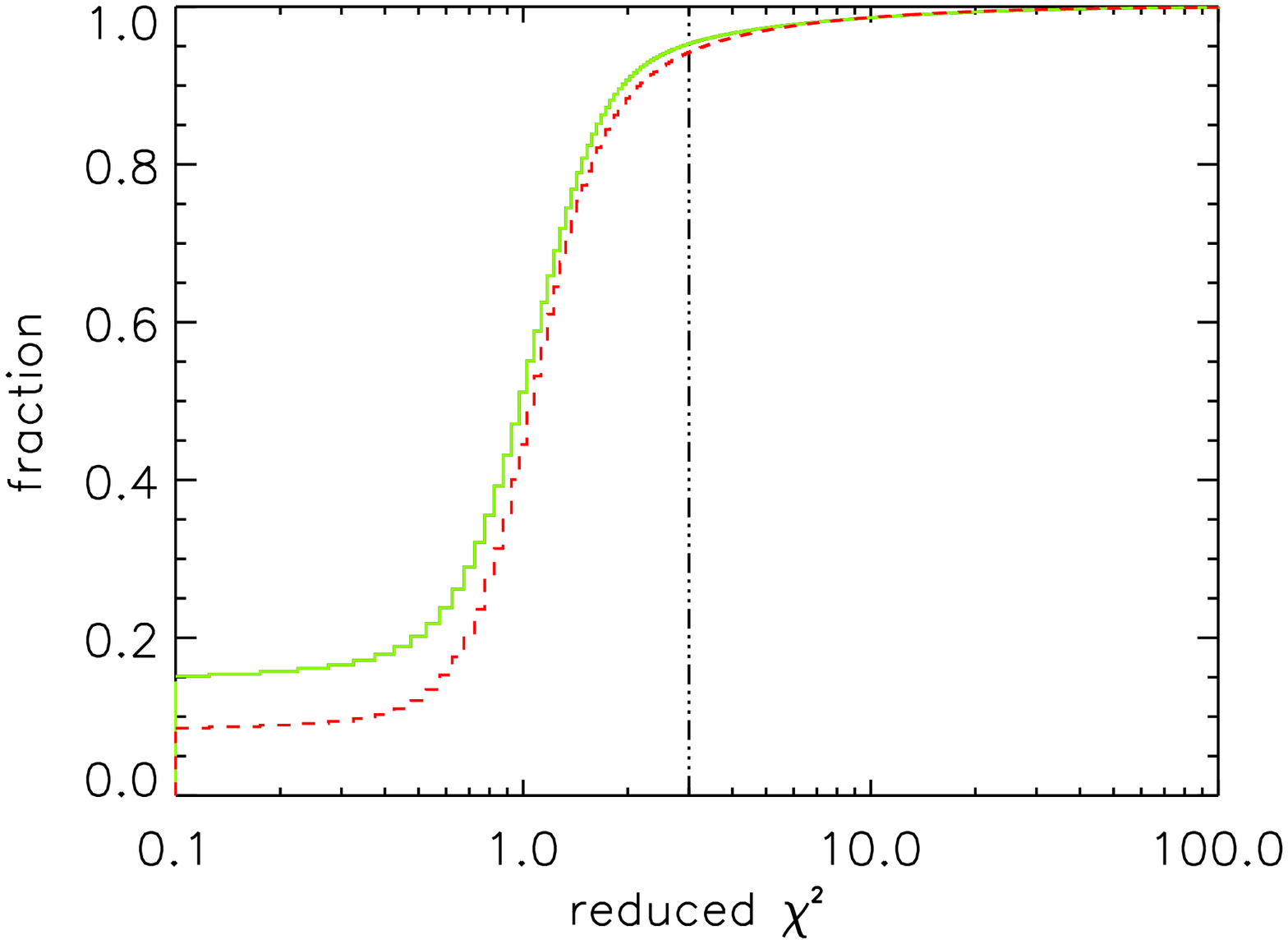,angle=0.0,width=0.4\linewidth}
\caption{Left: Cumulative distribution of mean object type in the
  HLC. Galaxy-like objects are assigned an object type of 3 and
  star-like objects are an object type of 6.  For multiple
  observations, the mean indicates whether the object is mostly
  classified as a star or a galaxy. The cut used to define the stellar
  sample is shown as a dashed vertical line.  Right: Cumulative
  distribution of reduced $\chi^2$ for $g$ (green, solid) and $r$
  (red, dashed) bands for all stellar objects. The slope of the
  distribution turns over at a reduced $\chi^2$ value of $\sim 3$,
  which is taken as the boundary separating constant and variable
  stars; this value is marked on the graph as a dash-dot line.}
\label{fig:mot}
\end{figure*}
\begin{figure}
\epsfig{file=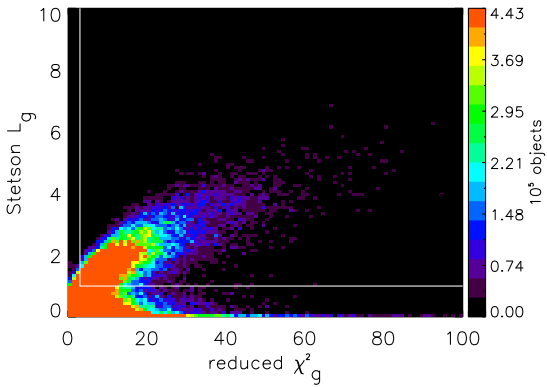,angle=0,width=0.9\linewidth}
\caption{Density plot of reduced $\chi^2_g$ against Stetson index
  $L_g$ for the stellar catalogue. Pixel colour represents the number
  of objects in each pixel bin. The cuts used to extract the high
  quality sample of stellar variables are shown as white lines.}
\label{fig:cuts}
\end{figure}

In Aristotle's model of the Universe, the stars were fixed on a
rotating sphere and eternally invariable. Stellar variability has been
known since Fabricius' discovery of Miras in 1596, although the
ancient Chinese and Korean astronomers were already familiar with
supernovae or ``guest stars''~\citep{Cl79}.  Proper motions were
discovered in 1718 by Edmund Halley, who noticed that Sirius, Arcturus
and Aldebaran had moved from their fixed positions recorded in
Aristotelian cosmology.

Despite this long history, our knowledge of both variable stars and
high proper motion sources remains very incomplete. As \citet{Pa00}
has emphasised, over 90 per cent of variable stars brighter than 12
mag have not been discovered.  There are still comparatively few large
archives of variable sources and, as a consequence, our knowledge of
many classes of object, including novae, supernovae, RR Lyraes (the
focus of this paper) and high proper motion objects remains
limited. Indeed, the variable sky remains one of the most unexplored
areas in astronomy, with the exciting possibility that even bright
variable objects may correspond to completely unknown astronomical
phenomena~\citep[see e.g.,][]{Pa01}.

The modern era of massive variability searches begins with the
microlensing surveys like MACHO \citep{Al93}, EROS \citep{Au93} and
OGLE \citep{Ud92}. Typically, these surveys monitored millions of
stars down to $V$$\sim$21 a few times every night over several years
in the directions of the Galactic Bulge and the Magellanic Clouds.
The resulting huge databases of lightcurves yielded information on
many rare types of astrophysical variability. They were the first
projects that harnessed the power of large format CCD cameras and
modern computers to show that the acquisition, processing and
archiving of millions of photometric measurements was feasible. The
surveys were soon followed by high-redshift supernova surveys (such as
High-Z SN Search \citep{Sc98} and the Supernova Cosmology project 
\citep{Pe99}), which typically had a lower time resolution and smaller area 
coverage than the microlensing surveys, although much deeper limiting 
magnitudes.

\begin{figure*}
\epsfig{file=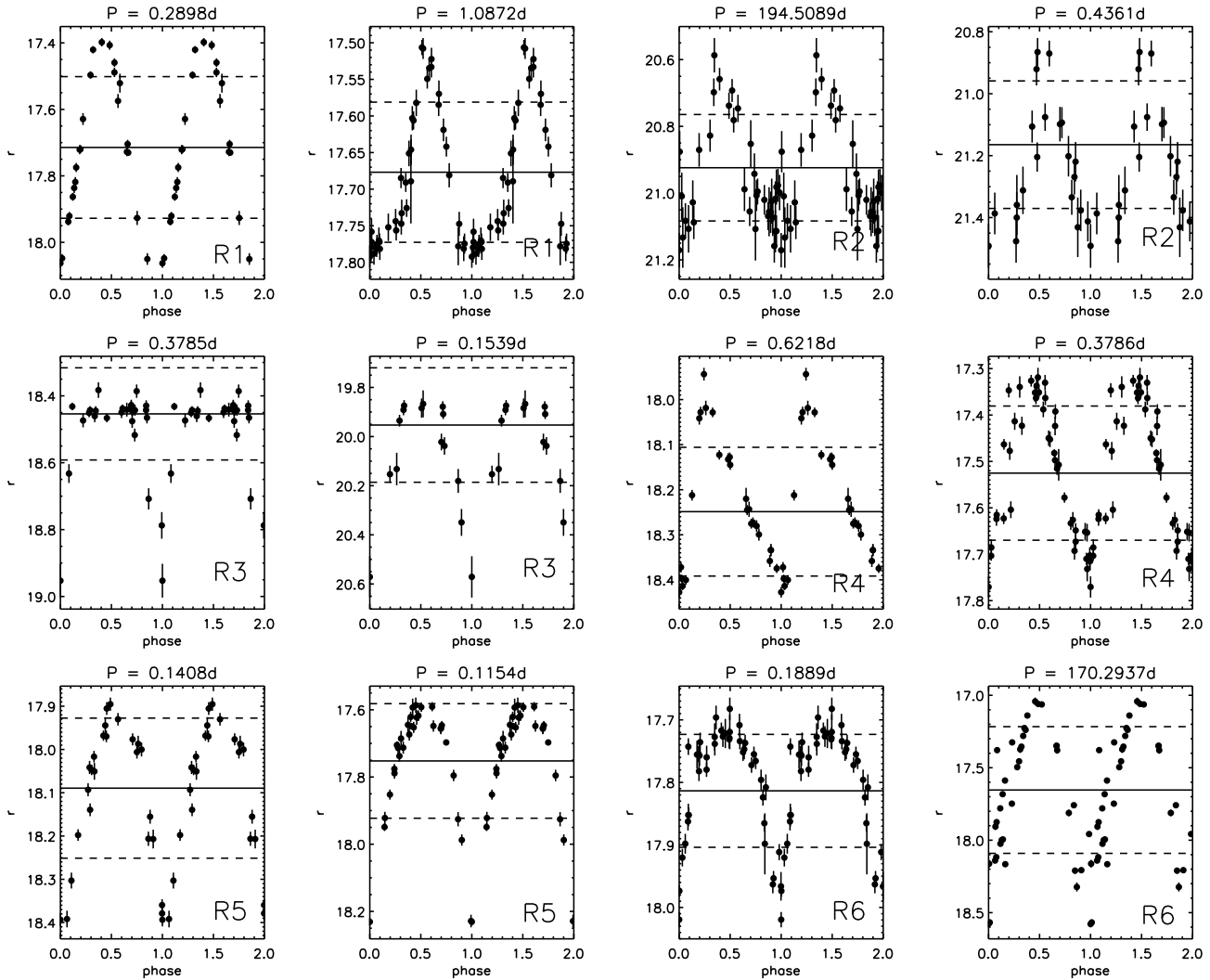,angle=0,width=\linewidth}
\caption{A sample of folded lightcurves. The period in days is
  recorded at the top of each plot. The solid line is the mean
  magnitude, whilst the dashed lines represent $1\sigma$
  deviations. In the bottom right corner, the number refers to the
  region in the colour-colour plot in which the lightcurve lies (see
  Figure~\ref{fig:colcol} in Appendix \ref{app:compsesar}). The two rightmost
  lightcurves in the middle row are probable RR Lyraes.}
\label{fig:lcs}
\end{figure*}

The Sloan Digital Sky Survey \citep[SDSS;][]{Yo00} provides deep and
homogeneous photometry in five bands in a large area around the North
Galactic Cap, but with almost no variability information.  The main
exception is the compilation of repeat scans of the $\sim$290
square degree area -- known as Stripe 82
\citep[see e.g.,][]{Ad08}.  The dataset has allowed the discovery of many new
supernovae, which are publicised and followed up spectroscopically with
other telescopes~\citep[see e.g.,][]{Fr08,Di08}.  
By averaging
repeat observations of unresolved sources in Stripe 82, \citet{Iv07}
built a catalogue of $1$ million standard stars with $r$ magnitudes
between $14$ and $22$. \citet{Se07} then carried out the first
analysis of $\sim$1.4 million variable stars and quasars
using a colour-colour plot to assign variable types.

Recently, \citet{Br08} presented a catalogue of almost four million
``light-motion curves'' using the data available in Stripe 82. Objects
are matched between the $\sim$30 epochs, taking into account the
effects of any proper motion over the eight-year baseline stretching
back to the earliest runs in 1998. Thanks to the high quality of the
SDSS imaging, excellent astrometric and photometric precision is
attainable.  \citet{Br08} also provide a Higher-Level Catalogue which
is a set of 229 derived-quantities for each light-motion-curve. These
quantities describe the mean magnitudes, photometric variability and
astrometric motion of a subset of objects whose light-motion-curve
entries pass certain quality constraints.

Variable stars of particular interest are RR Lyraes, which have often
been used to identify Galactic substructure -- for example, in studies
of the Sagittarius Stream~\citep{Iv00,Vi04,Ke08}, and the Monoceros
and Virgo Overdensities~\citep{Vi06,Ke09}. RR Lyraes are particularly
useful for three reasons.  First, they are constituents of the old,
metal-poor halo, in which substructure is abundant. Second, they are
standard candles, enabling an estimate of their distances to be
made. Third, they are bright enough to be detected out to distances of
$\sim$130\,kpc in the SDSS data, giving us an insight into the
structure of the remote Milky Way halo.

In this paper, we select first the variable objects in Stripe 82 and
then the subset of RR Lyraes, using the \citet{Br08} Light-Motion
Curve Catalogue (LMCC) and Higher-Level Catalogue (HLC). \S
\ref{sect:vars} discusses the selection of the variable objects and
their properties, whilst \S \ref{sect:rrlyraes} describes the
identification of the RR Lyrae stars and properties of the population
in some detail. \S \ref{sect:lyr_subst} discusses substructure seen in
the distribution of the RR Lyraes and we conclude in \S
\ref{sect:concl}.

\section{Variable Stars in Stripe 82}
\label{sect:vars}

\subsection{Variable Selection}
\label{sect:varselect}

\begin{figure*}
\epsfig{file=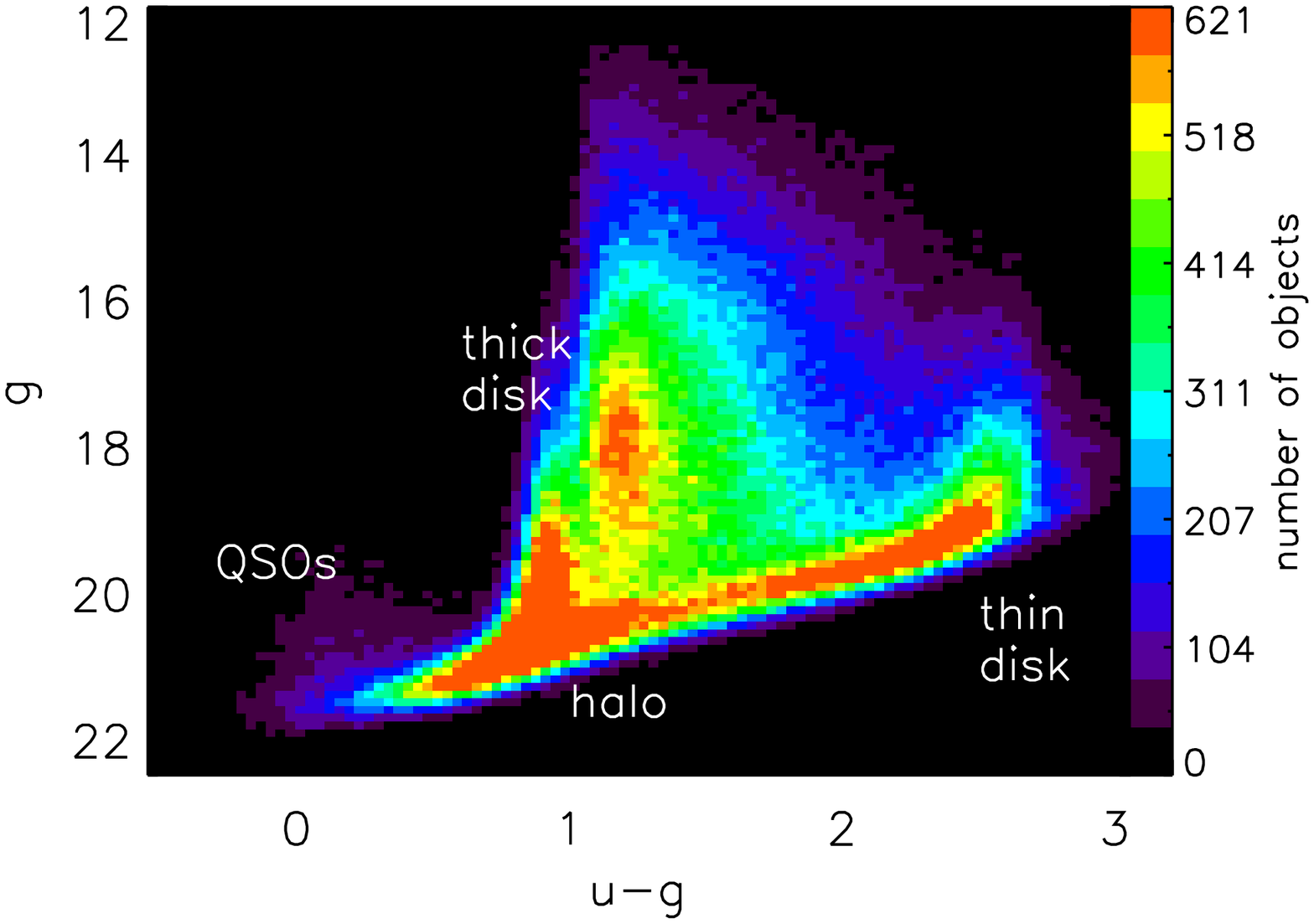,angle=0,width=0.45\linewidth}
\epsfig{file=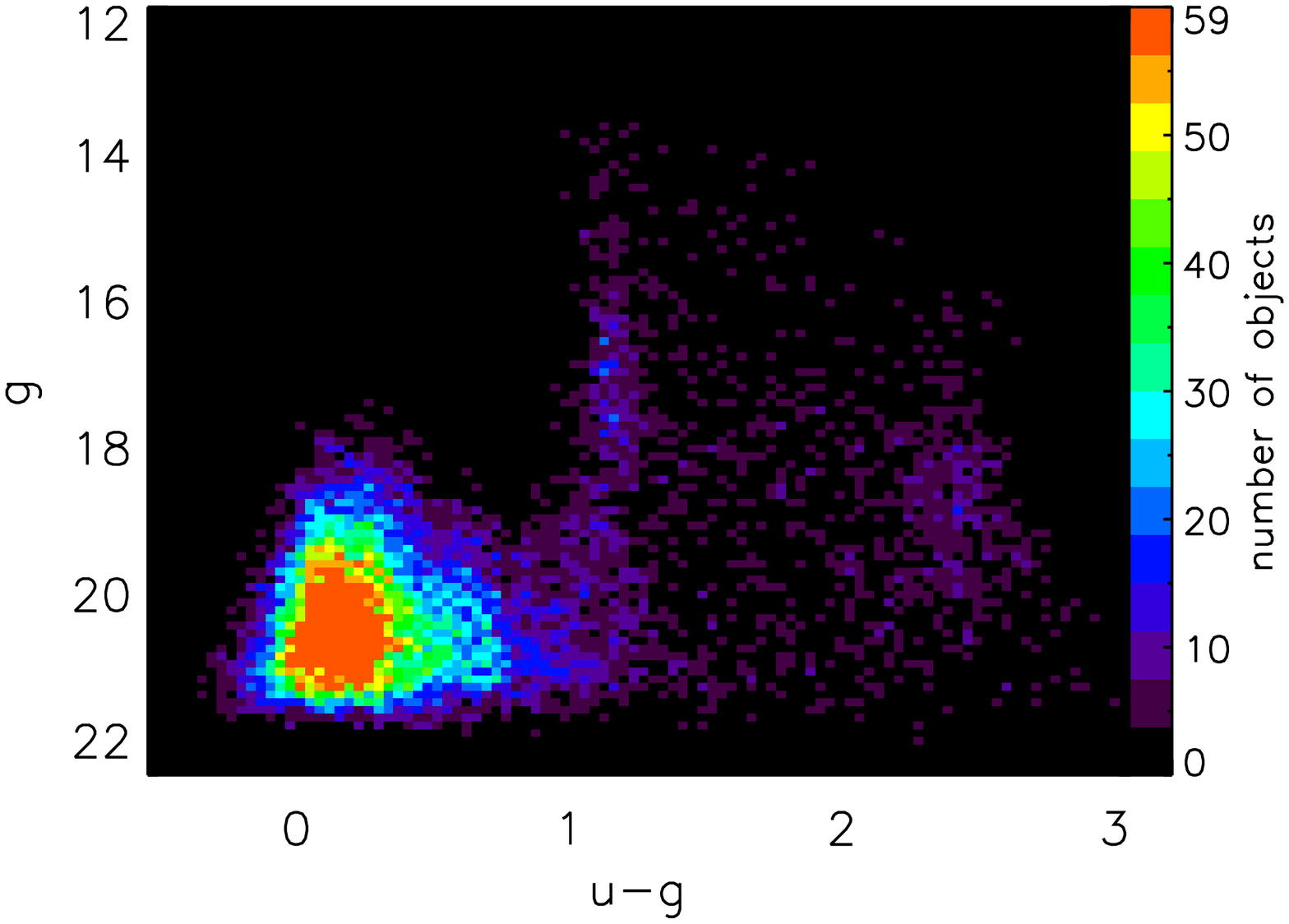,angle=0,width=0.45\linewidth}
\caption{Left: Colour-magnitude diagram ($g$ versus $u-g$) for all
  527\,621 sources in the stellar sample for which high-quality $u$
  and $g$ data exist.  Right: Colour-magnitude diagram ($g$ versus
  $u-g$) for the subset of 21\,939 variables.}
\label{fig:cmds}
\end{figure*}

The HLC contains 3,700,548 objects.  Every object in the survey is
assigned an object type each time it is observed: 3 if it is
galaxy-like and 6 if it is star-like \citep[see \S~3.1 of][]{Br08}.
For multi-epoch data, the object type averaged over all the
observations provides a relatively reliable indicator of whether the
source is star-like or galaxy-like.  The cumulative distribution of
mean object type in the left panel of Figure~\ref{fig:mot} shows that
$\sim 55$ per cent of objects in the catalogue are purely star-like or
purely galaxy-like.  To extract a sample of stars with essentially
zero contamination from galaxies, we require that the mean object type
is 5.5 or greater. This results in a ``stellar'' sample of 1\,671\,582
objects.

For the stellar sample, a cumulative distribution of reduced $\chi^2$
(that is, $\chi^2$ per degree of freedom) for the $g$ and $r$ bands
is shown in the right panel of Figure~\ref{fig:mot}. The value of
reduced $\chi^2$ at which the distributions turn over is $\sim$3,
which is taken as the $\chi^2$-value below which stars are assumed to be
well-modelled by a constant baseline. The number of objects that
simultaneously satisfy reduced $\chi^2 >3$ in both $g$ and $r$ is
41\,729.

The stars with reduced $\chi^2 > 3$ in both $g$ and $r$ bands are
mainly variables, but still contain some artifacts, typically due to
one or two outlying photometric measurements. One way to test for true
variability is to look for correlations between different bands; a
true variable star will usually have changes in brightness that are
correlated in all bands whereas a discordant measurement may exist in
one band only. Throughout this analysis, the Stetson index ($L_g$) is
used as a measure for correlated variability between the $g$ and $r$
band data (see Stetson 1996).

For the stellar sample, a density plot of reduced $\chi^2$ in the $g$
band against the Stetson $L_g$ is shown in Figure~\ref{fig:cuts}, in
which two distinct populations can be discerned. The first has an
almost linear correlation between reduced $\chi^2$ and $L_g$, and are
almost all true variable stars.  The second has a high reduced
$\chi^2$, but $L_g$ is low, indicating that the brightness changes
which give rise to the high reduced $\chi^2$ values are not correlated
between bands. To extract a sample of high-quality variable stars, we
impose the simultaneous restrictions $L_g >1$, reduced $\chi^2_r >3$
and reduced $\chi^2_g >3$, together with requiring at least 10 good
epochs \citep[see][]{Br08}, leaving 21\,939 objects.  We present a
comparison of the content of our variable catalogue with the earlier
catalogue of \citet{Se07} in Appendix \ref{app:compsesar}.

To show the quality of the data, a selection of folded light-motion
curves are shown in Figure~\ref{fig:lcs}, from which a variety of
periodic phenomena such as stellar variability and eclipses are
evident. In particular, the two right-most images in the middle row
are very likely RR Lyraes. From the periods and the lightcurve shapes,
we might surmise that the first lightcurve is an ab-type RR Lyrae and
the second is a c-type RR Lyrae.

\subsection{Variable Properties}
\label{sect:varprops}

Colour-magnitude diagrams ($g$ versus $u-g$) are plotted for both the
stellar sample with $u$ band data and the subset of variable stars, in
Figure~\ref{fig:cmds}. For the stellar sample, we see three prominent
clumps associated with the thin disk, thick disk and halo on moving
redwards in colour. For the variables, there is one prominent clump
centered on $u-g \approx 0.2$, which is primarily associated with
variable quasars. There are less prominent, but still discernible,
peaks associated with variable stars in the thin disk, thick disk and
halo.

A crude discrimination between different classes of variable objects
is possible in $g-r$ versus $u-g$ space.  We find that the variable
sample is largely comprised of stellar locus stars and low-redshift
quasars; stellar locus stars are predominant at bright ($g < 19$)
magnitudes while low-redshift quasars dominate at faint ($g < 22$)
magnitudes.  RR Lyraes also make a significant contribution (see
Appendix~\ref{app:compsesar} for more detail on the statistical
properties of the variable sample).

\subsection{The Proper Motions}

\begin{figure}
\epsfig{file=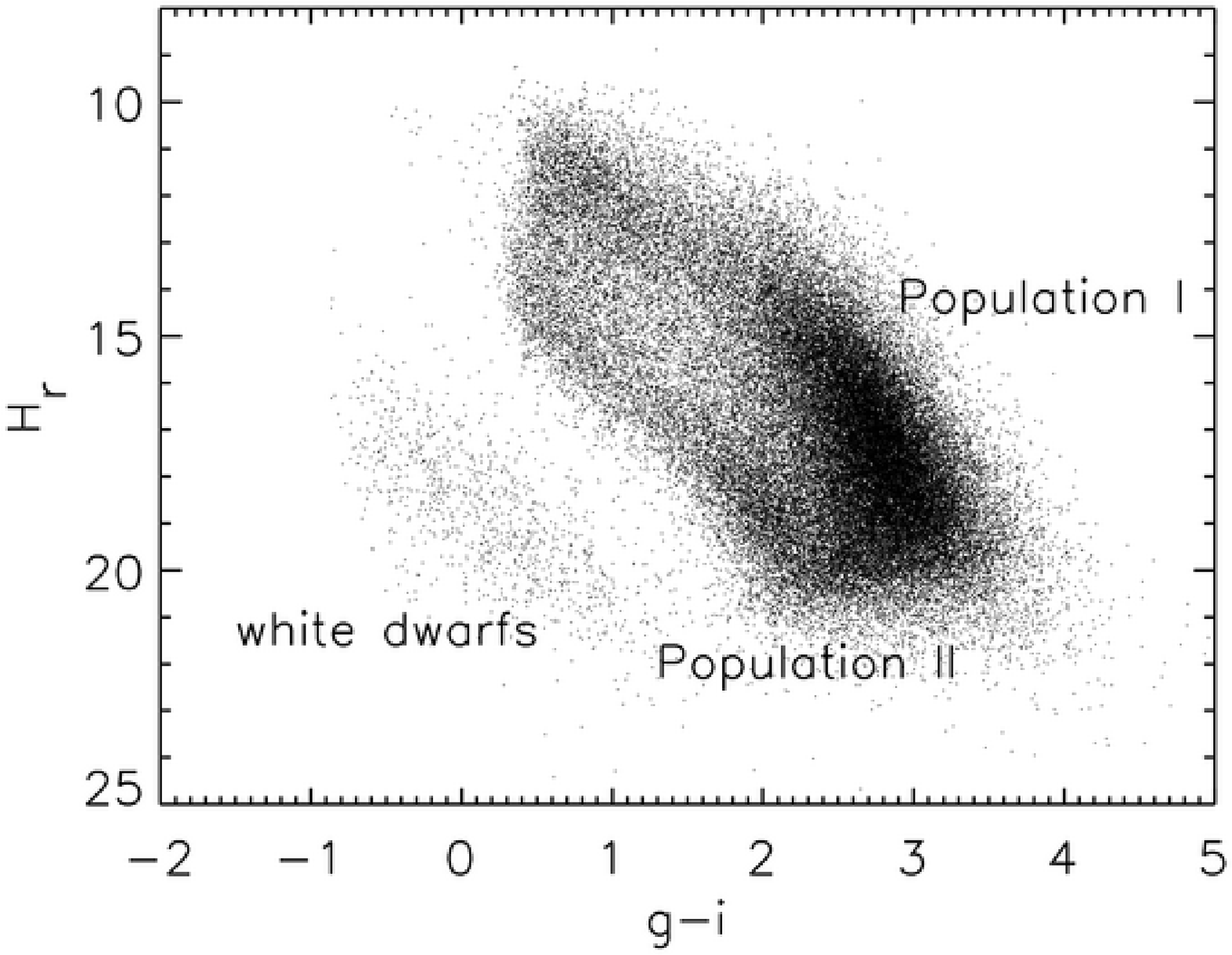,width=0.9\linewidth}
\epsfig{file=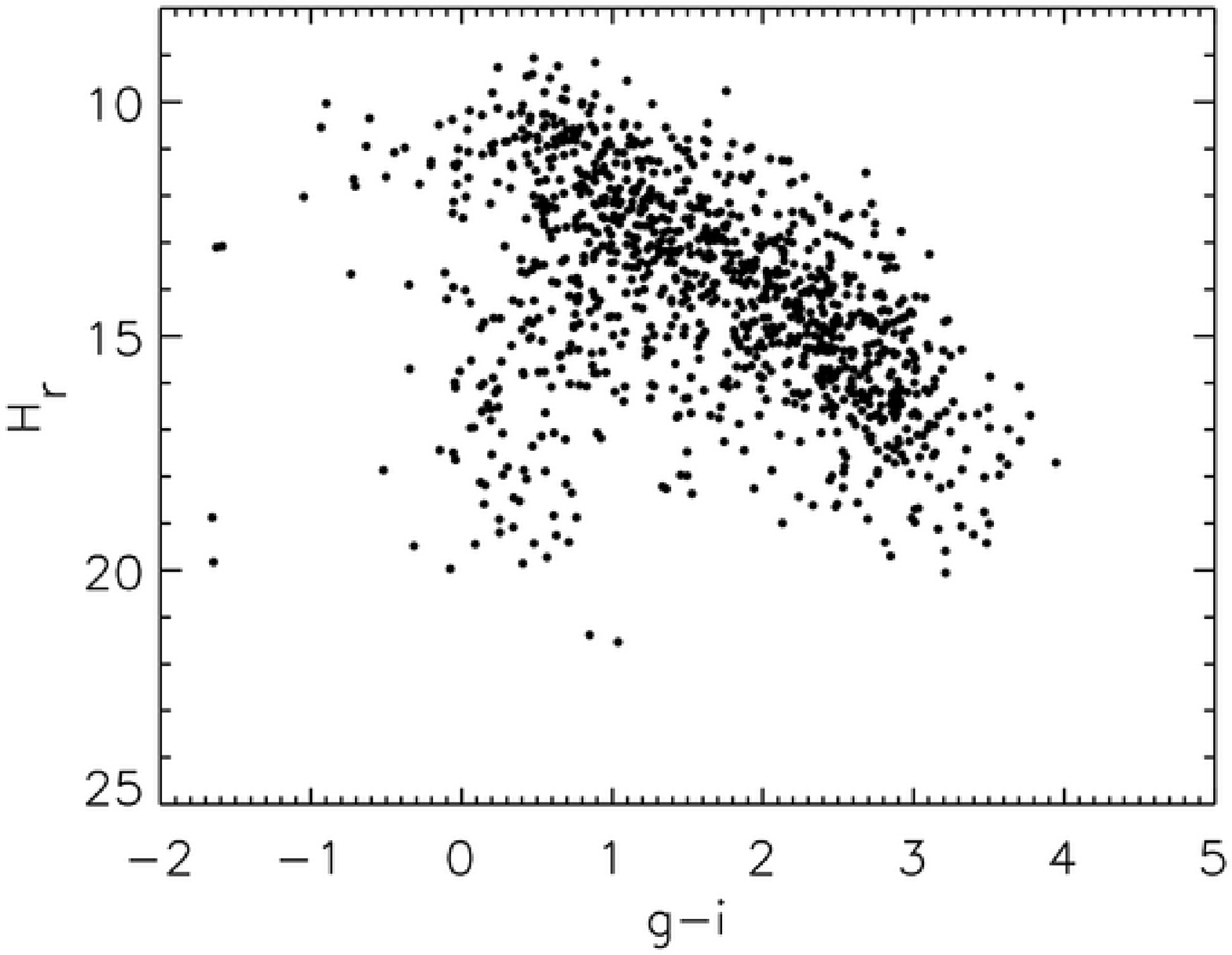,angle=0,width=0.9\linewidth}
\caption{Top: Reduced proper motion diagram for all objects in the HLC
  with proper motion of S/N $>10$ and $|\mu| > 2$ mas\,yr$^{-1}$.
  Bottom: Reduced proper motion diagram for all objects in the
  variable subset with S/N $>5$ and $|\mu| > 2$
  mas\,yr$^{-1}$.}
\label{fig:rpms}
\end{figure}

The HLC offers an improvement over previous variability work in Stripe 82 
through the availability of proper motions. The combination of stellar 
photometry and proper motions has proved to be a powerful way of classifying 
stars -- in particular, members of elusive populations such as white dwarfs, 
brown dwarfs and wide binaries. Such combined catalogues, drawn from the 
intersections of SDSS data with USNO-B data~\citep{Mo03}, have been 
constructed before~\citep{Mu04,Go04,Ki06}. Compared to such catalogues, the 
HLC is restricted to Stripe 82 and the proper motion sensitivity is poorer, 
due to the much shorter time baseline.  On the other hand, the Stripe 82 
photometric catalogue extends approximately $1.5$ magnitudes deeper than the 
limiting magnitude of USNO-B proper motions ($V$$\sim$21).

The reduced proper motion is defined as $H = r + 5 \log \mu +5$, where $r$ is 
the apparent magnitude and $\mu$ the proper motion in arcsec\,yr$^{-1}$. The 
criteria for inclusion in the reduced proper motion diagram in the top panel 
of Figure~\ref{fig:rpms} is that the object lies in our stellar sample, that 
the proper motion is measured with a signal-to-noise ratio (S/N) $>$10 and 
that the absolute value of the proper motion $|\mu|$ exceeds 
2\,mas\,yr$^{-1}$. The S/N-cut has been chosen primarily to ensure easy 
visibility of structure on the figure, whilst the proper motion cut enables us 
to excise quasars. We discern three distinct sequences of stars, namely 
Population I disk dwarfs, Population II main sequence subdwarfs and disk white 
dwarfs. \citet{Vi07} have used this reduced proper motion diagram to identify 
new ultra-cool and halo white dwarfs in Stripe 82, Similarly, 
\citet{Sm08a,Sm08b} have used the same procedure to extract a sample of halo 
subdwarfs in studies of the velocity ellipsoid and halo substructure.  The 
lower panel of Figure~\ref{fig:rpms} shows the reduced proper motion diagram 
just for the subset of variables with proper motion S/N$>$5 and 
$|\mu|$$>$2\,mas\,yr$^{-1}$.  The variables are disproportionately drawn from 
the Population I disk dwarfs, although the other two sequences can still be 
seen.

\section{RR Lyraes}
\label{sect:rrlyraes}

\subsection{Identification of RR Lyraes}
\label{sect:rrlyraes_extraction}

Here, we develop the tools to extract a high quality sample of RR
Lyraes from the variable catalogue. We begin by selecting the 873
candidates that simultaneously satisfy all the following criteria,
which are adapted from \citet{Iv05}, namely
\begin{eqnarray}
13.5 < r < 20.7, & \qquad & 0.98 < u-g < 1.35, \nonumber\\
-0.16 < r-i < 0.22,&\qquad & -0.21 < i-z < 0.25, \\
D^{min}_{ug} < D_{ug} < 0.35, &\qquad & D^{min}_{gr} < D_{gr} < 0.55,\nonumber
\end{eqnarray}
where
\begin{equation}
D_{ug} = (u\!-\!g)+0.67(g\!-\!r)\!-\!1.07,
\quad D_{gr}=0.45(u\!-\!g)\!-\!(g\!-\!r)\!-\!0.12.
\label{eqn:rrl_dug}
\end{equation}
The $r$ band magnitudes correspond to distances $\sim$5-130\,kpc.
$D_{ug}$ and $D_{gr}$ represent slopes in the $u-g$ and $g-r$
colour-colour plane. Combined with the cut on $u-g$, the $D_{ug}$ and
$D_{gr}$ criteria constrain the RR Lyraes to a hexagonal box in
colour-colour space, optimizing the selection of RR Lyraes.  The
values $D^{min}_{ug}$ and $D^{min}_{gr}$ can be altered to adjust the
completeness and efficiency of the RR Lyrae selection.  We chose to
use values $D^{min}_{ug} =$ -0.05 and $D^{min}_{gr} =$ 0.06, which
would give a completeness of 100 per cent for the QUEST survey RR
Lyraes~\citep{Vi04,Iv05}.

\subsection{RR Lyrae Periods}
\label{sect:rrlyrae_periods}

Determining periods for our RR Lyrae candidates is non-trivial. In
general, there are 30 to 40 datapoints in a lightcurve, unevenly
sampled over an eight-year baseline.  From this sparsely-sampled data,
we seek a period that is a fraction of a day.  The multi-band nature
of the SDSS survey is an advantage here, as we are able to verify that
any period estimate we obtain in one band is consistent with the data
in additional bands.

The LMCC contains all the data for a given lightcurve.  Each datapoint
has a flag that is set (or unset) if the datapoint passes (or fails)
certain quality requirements \citep[for more details, see][]{Br08}.
For period estimation, it is important that we use only the reliable
data to minimise errors.  In general, the $g$ and $r$ bands have
smaller errors and fewer outliers and their clean lightcurves are the
best sampled, so these bands are used together to estimate periods for
our lightcurves.

\begin{figure}
\epsfig{file=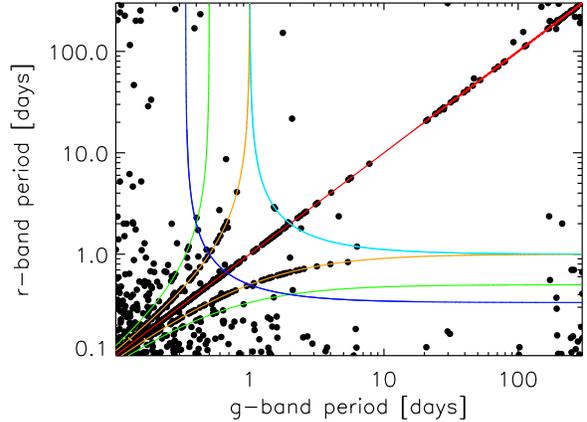,angle=0,width=\linewidth}
\caption{Lomb-Scargle period estimates in the $g$ and $r$ bands for
  the candidate RR Lyraes. The coloured lines represent resonance
  lines along which one period is a harmonic of the other: $P_r = P_g$
  (red), $P_r = P_g/(1 \pm P_g)$ (orange), $P_r = (P_g/1 \pm 2P_g)$
  (green), $P_r = P_g/(P_g-1)$ (cyan), $P_r = P_g/(2P_g-1)$ (blue).}
\label{fig:lyr_pgpr}
\end{figure}

As a first pass, we run a Lomb-Scargle periodogram~\citep[see
e.g.,][]{NR5} on each of the $g$ and $r$ band lightcurves, taking care
to ensure that the frequency range extends to the high frequencies (or
low periods) expected for RR Lyrae stars and that the sampling rate is
detailed enough to discern individual peaks.  The resulting period
estimates are plotted in Figure~\ref{fig:lyr_pgpr}.  First, we note
that there are a number of candidates which have matching period
estimates.  A number of resonance lines -- the locations at which one
of the period estimates is a harmonic of the other -- are plotted as
solid lines in the graph.

\begin{figure}
\epsfig{file=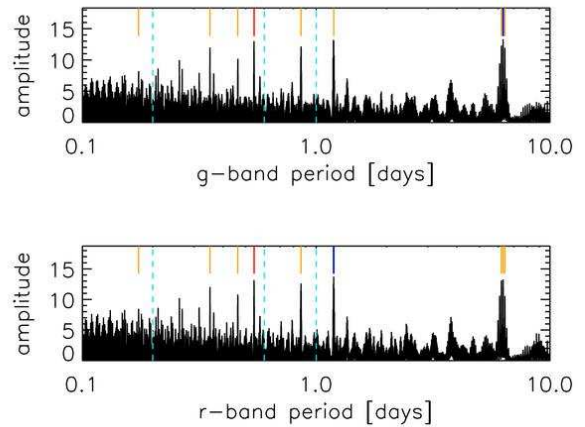,angle=0,width=\linewidth}
\caption{A sample periodogram spectrum.  The highest peak in each
  spectrum is indicated by the blue line; the additional peaks
  selected by the method described in the text are indicated by the
  orange lines, with the eventual best-fit peak marked with a red
  line.  At least one peak was found in each region delineated by the
  cyan lines.}
\label{fig:pdgram}
\end{figure}
\begin{figure}
\epsfig{file=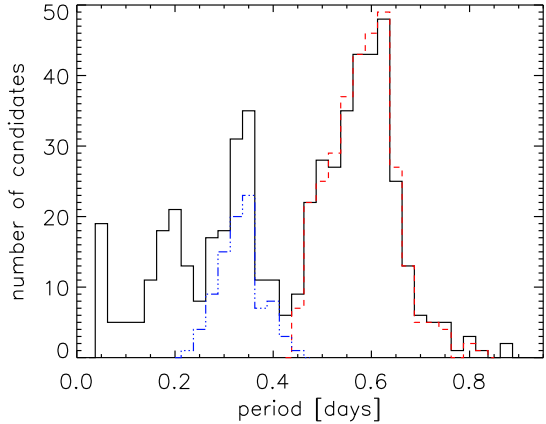,angle=0,width=\linewidth}
\caption{The distribution of periods for all RR Lyrae candidates with
  periods in the expected range for RR Lyraes (black, solid), for the
  ab-type RR Lyraes (red, dashed) and the c-type RR Lyraes (blue,
  dot-dashed).  The spike at $P<0.1$ is due in part to $\delta$ Scuti
  and SX Phe stars, and the population with the shortest periods is due mostly
  to eclipsing variables.}
\label{fig:lyr_pdistbn}
\end{figure}
The resonance lines are well-populated, indicating that the
Lomb-Scargle periodogram can return harmonics of the true period as
well as the period itself.  Hence, we must consider whether the exact
period matches are indeed cases where the true period has been
recovered, or whether both period estimates are harmonics, and so
on. Lightcurves with periods that do not match or do not lie on
resonance lines could be objects that are not periodically variable --
quasars, for example -- or objects for which the periodogram has
failed to recover the true period or a harmonic of the period in one
or both cases. Not all of the period estimates lie within the range
expected for RR Lyraes ($0.2-0.8$ days), which is probably a
consequence of contamination in our sample.  However, it is unclear
whether we can simply reject any objects with period estimates larger
than those expected for RR Lyraes.  Certainly, some of the larger
period estimates could be attributed to the sparse sampling of the
lightcurves, which may generate a signal that overwhelms the periodic
nature of the RR Lyraes.

Accordingly, we experimented with a number of alternative methods,
including binning, smoothing and phase dispersion minimization, before
concluding that a string-length technique is the most effective for
our dataset.  In outline, a string-length method works by phasing a
lightcurve with a trial period and calculating the sum of the
straight-line distances between consecutive points.  The sum of these
distances, or the string-length, will be minimised if the trial period
is the true period.

We use a variation of the \citet{La65} string-length technique
described in \citet{St96}.  For each trial period, the string-length
is computed for the $g$ and $r$ bands. Their sum is taken as the
overall string-length, which is minimised to obtain a period
estimate. Running a string-length period finder over a wide and
finely-sampled range of trial periods is computationally expensive.
However, we can short-cut the process by restricting the string-length
period search to a narrow range of periods, centred on a set
of the most likely periods identified via the Lombe-Scargle periodogram.

To ensure that the correct RR Lyrae periods are identified, the $g$
and $r$ band Lombe-Scargle periodogram spectra were combined and the
highest peaks in each of the ranges $P$$<$0.2, 0.2$<$$P$$<$0.6, 
0.6$<$$P$$<$1.0 and $P$$>$1.0\,days were selected. Then, four further peaks
were selected from each of the $g$ and $r$ band spectra
independently. This was done according to highest amplitude, until four
distinct new peaks had been selected in each band.  Each peak was then
considered in turn: the string-length was calculated for a narrow
range of periods spanning the peak, the period for which the
string-length was minimised was taken to be the best period in the
vicinity of that peak.  Finally, the period that returned the shortest
string-length overall was adopted as the period estimate for the
lightcurve.

In Figure~\ref{fig:lyr_pdistbn}, the solid black line is the period
distribution for all of the RR Lyrae candidates and shows four clear
peaks.  Moving from high to low periods, these populations are
predominantly: ab-type RR Lyraes (peak at $\sim$0.6\,d), c-type RR
Lyraes (peak at $\sim$0.35\,d), eclipses (peak at $\sim$0.18\,d) and
$\delta$ Scuti and SX Phe stars (peak at $\sim$0.05\,d).  Also present
in this candidate sample, are a number of non-periodic variables.  The
red dashed line is the period distribution for only those stars later
determined to be ab-type RR Lyraes and the blue dot-dashed line is for
those stars we later determine to be c-type RR Lyraes.

Before proceeding any further, we clean the sample of some eclipsing
variable stars, $\delta$ Scuti stars, SX Phe stars and non-periodic
variable contaminants by adopting a stringent cut on Stetson index
$L_g$$>$2.5. To perform subsequent analysis, we require that objects
have a sufficient number of clean data points; that is, we impose
further cuts on the number of clean epochs in the $g$ and $r$ bands:
$N_g$$>$5 and $N_r$$>$5, leaving 604 candidates.

Uncertainties in the period estimates may be attributed to two
sources.  If we assume that the estimate is indeed close to the true
period and not a harmonic, then any error is due to the string-length
fitting technique, estimated to be $\sim$10$^{-5}$\,day, from analysis
of the phased RR Lyrae lightcurves. However, for a small fraction of
the sample, where the period estimate is a harmonic of the true
period, the error will be $\sim$0.1-0.5\,day.

\subsection{RR Lyrae Classification}
\label{sect:lyr_cleansample}
\begin{figure}
  \epsfig{file=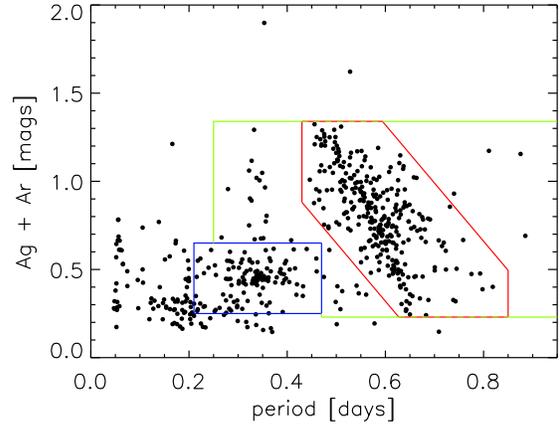,angle=0,width=\linewidth}
  \caption{Period versus the combined $g$ and $r$ band
    amplitudes. Three selection boxes are shown: red for candidate
    RRabs, blue for candidate RRcs, and green for candidates
    warranting further study.}
\label{fig:lyr_pdamp_cands}
\end{figure}
\begin{figure}
  \epsfig{file=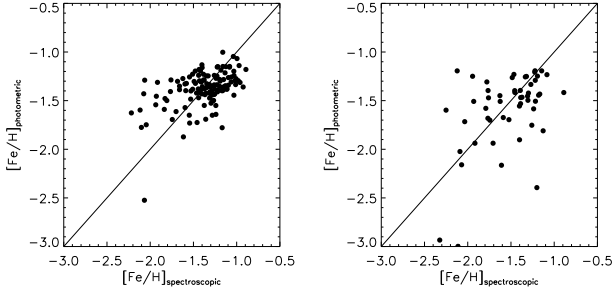,angle=0,width=\linewidth}
  \caption{The performance of the photometric metallicity
    relationships in eqns~(\ref{eq:rrab_metrln}) and
    (\ref{eq:rrc_metrln}) is shown. The left (right) panel plots the
    RRab (RRc) candidates with spectroscopic metallicities against the
    corresponding photometric estimate.}
\label{fig:calib}
\end{figure}

\citet{Pi02} first divided RR Lyraes into three classes - a, b and c -
based on the appearance of their lightcurves.  Further study of Lyraes
has altered the classification instead to just ab and c.  It is
believed that RRab stars are pulsating in the fundamental mode and RRc
stars are pulsating in their first overtone~\citep[e.g.,][]{Sm95}.

The two classes have somewhat different properties: RRab stars
generally have relatively large amplitudes (of order a magnitude) and
asymmetric lightcurves with a steep rising branch and a slow, steady
decline.  Their periods lie mostly in the range 0.4-1.0\,day and
studies of RR Lyraes in globular clusters have shown that the
amplitude of RRab stars decreases with increasing period.  RRc stars
have smaller visual amplitudes (around half a magnitude) and their
lightcurves are more symmetrical and nearly sinusoidal.  They
generally have periods of 0.24-0.5\,day.

To separate the two classes, we proceed by plotting the candidates in
the plane of $(P,A)$, where $P$ is the period and $A$ is the sum of
the amplitudes in the $g$ and $r$ band lightcurves, as shown in
Figure~\ref{fig:lyr_pdamp_cands}.  The red box is defined as the
region
\begin{eqnarray}
& & 0.43 < P < 0.85, \qquad 0.23 < A < 1.34 \nonumber \\
& & 2.3 < A + 3.3P < 3.3
\end{eqnarray}
and includes 296 RRab candidates. The blue box is defined as the region
\begin{equation}
0.21 < P < 0.47, \qquad 0.25 < A< 0.65
\end{equation}
and includes 122 RRc candidates.

Finally, the green boxes (not fully shown in Figure~\ref{fig:lyr_pdamp_cands} 
for clarity, but which extend to a period of two days) include a number of 
further candidates, which we do not want to discard without further 
investigation.  Objects with a period of one day are almost certainly spurious 
(the lightcurves have a sampling period of one day) and are removed.

For the 43 objects that remain, there is the possibility that they may lie 
away from the concentration of RR Lyraes because of an incorrect period 
estimate. As discussed in Section \ref{sect:rrlyrae_periods}, errors in the 
period estimates are small, but sometimes a harmonic of the true period is 
obtained. Hence, we revisit the period analysis to determine whether any of 
the likely period peaks lie within the red RRab box or the blue RRc box.  If 
more than one period peak lies within a box, the period with the minimum 
string-length is used.  Any objects for which a fitting period can be found 
are added to the RRab or the RRc candidate set as appropriate. This 
reclassification results in 330 RRab and 137 RRc candidates.

\subsection{The RRab types}
\label{sect:rrlyraes_rrabs}

126 of our RRab and 52 RRc candidates possess SDSS spectra and have 
spectroscopic metallicity estimates. We use these objects to calibrate 
empirical relationships and thence derive photometric metallicities for the 
entire sample.

\begin{figure}
  \epsfig{file=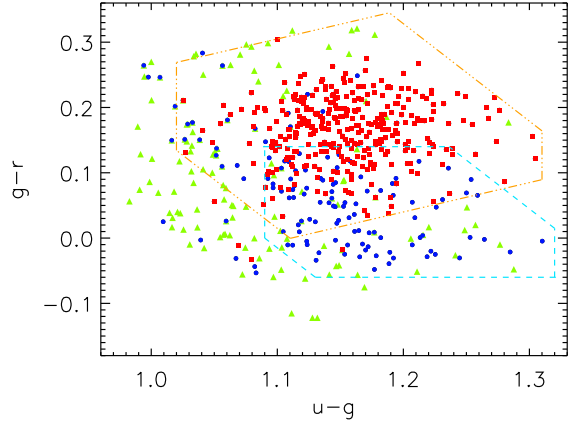,angle=0,width=\linewidth}
  \caption{The RR Lyrae candidate selection boxes in the $u\!-\!g$ versus 
    $g\!-\!r$ plane; the orange (dot-dashed) box is used to select RRab types, 
    the cyan (dashed) box for RRc types. The RRab (RRc) candidates that pass 
    the period, amplitude and metallicity cuts are shown as red squares (blue 
    circles). Also shown are suspected eclipsing variables as green triangles.}
  \label{fig:lyr_cands_grug}
\end{figure}

\cite{Ju96} found that the metallicity of an RR Lyrae depends on the period 
$P$ and the shape of the lightcurve, which may be parameterised via a Fourier 
decomposition:
\begin{equation}
	f(\theta) = A_0 + \sum_{i=1}^{N} A_i \sin(i\theta + \phi_i).
	\label{eqn:fouriersin}
\end{equation}
The amplitudes $A_i$ and the phases $\phi_i$ can then be combined as follows:
\begin{equation}
	A_{ij} = \frac{A_i}{A_j},\qquad\qquad
	\phi_{ij} = j\phi_i - i\phi_j.
	\label{eqn:phiij}
\end{equation}
Inspired by the analogous relation of \cite{Ju96}, we use the spectroscopic 
metallicities and the lightcurve properties to derive the metallicity-period-
amplitude-phase relation
\begin{eqnarray}
  [{\rm Fe/H} ] & = & 0.845 - 4.487 P - 0.187 \phi_{31} - 1.454 A_{31} + 2.166 P \phi_{31} \nonumber \\ & & + 1.563 P A_{31} - 8.374 P^2 - 0.081 \phi_{31}^2.
	\label{eq:rrab_metrln}
\end{eqnarray}
which has a typical scatter $\sigma = 0.25$.  Its performance is shown in the 
left panel of Fig.~\ref{fig:calib}. To refine our RRab sample, we insist that
\begin{equation}
	-3 < \textrm{[Fe/H]} < 0,
\end{equation}
and apply a further restriction to trim the sample of a remaining few 
eclipsing variables by imposing the selection box in orange (dot-dashed lines) 
in colour-colour space shown in the lower panel of 
Figure~\ref{fig:lyr_cands_grug}.  This leaves us with 325 RRab candidates.  
Also shown in this figure are the confirmed RRab types as red squares and 
likely eclipsing variables as green triangles.

\subsection{The RRc types}
\label{sect:rrlyraes_rrcs}

\cite{Mo07} found that the metallicities of c-type RR Lyraes also vary with 
period and lightcurve shape.  Owing to the difference in pulsation mode, and 
hence lightcurve shape, between the RRab stars and the RRc stars, a new 
metallicity relation must be derived for the RRc candidates.  The lightcurves 
were Fourier decomposed and the 52 RRcs with spectroscopic metallicities were 
used to derive the following metallicity-period-amplitude relation, the form 
of which is suggested by ~\cite{Mo07}:
\begin{eqnarray}
  [{\rm Fe/H} ] & = & -10.669 + 60.944 P + 4.351 A_{21} \nonumber 
  \\ & & - 23.418 P A_{21} - 95.344 P^2 + 1.864 A_{21}^2.
  \label{eq:rrc_metrln}
\end{eqnarray}
As the right panel of Fig.~\ref{fig:calib} shows, there is a somewhat
greater scatter of $\sigma = 0.38$ in this relationship than the
corresponding one for RRabs.

A metallicity cut $-3.75 < \textrm{[Fe/H]} < 0$ removes a few obvious 
outliers. As for the RRabs, we discard eclipsing variables by imposing the 
selection box in cyan (dashed lines) in colour-colour space shown in the lower 
panel of Figure~\ref{fig:lyr_cands_grug}. This leaves us with 97 RRc type 
candidates. Also shown in this figure are the confirmed RRc types as blue dots 
and likely eclipsing variables as green triangles.

In the same way that the outlying candidates were reanalysed to determine 
whether a period could be found that placed the star into the RRab or RRc 
dataclouds, so the RRab and RRc rejects are also reanalysed, just in case they 
are misclassified RRc and RRab respectively.  The metallicity and colour cuts 
described above then determine inclusion in the candidate sets. Finally, a 
visual inspection of the lightcurves confirms that these are very high-quality 
samples. Judging by lightcurve shape, there are at most 21 possible 
contaminants in the RRab set and only 6 possible contaminants in the RRc set. 
The lightcurves typically possess low S/N and/or are poorly sampled. While 
many of these objects will be RR Lyrae, to be conservative, they are removed, 
giving us a final sample of 316 RRab stars and 91 RRc stars, 407 RR Lyraes in 
all.

\begin{table}
\centering
\begin{tabular}{l|rr|rr}
\hline
property & $\mu_{ab}$ & $\sigma_{ab}$ & $\mu_c$ & $\sigma_c$ \\
\hline
$A_g$ (mag) & 0.47 & 0.14 & 0.26 & 0.05 \\
$A_r$ (mag) & 0.34 & 0.11 & 0.19 & 0.03 \\
P (days) & 0.58 & 0.07 & 0.33 & 0.04 \\
$\chi^2_g$ & 363.8 & 326.1 & 146.6 & 80.0 \\
$\chi^2_r$ & 250.9 & 232.3 &  98.5 & 53.8 \\
$L_g$ & 10.7 & 5.2 & 8.4 & 2.5 \\
$[$Fe/H$]$ & -1.38 & 0.16 & -1.61 & 0.42 \\
$D$ (kpc) & 28.9 & 21.9 & 21.1 & 17.6 \\
$r$ (kpc) & 29.2 & 21.7 & 21.5 & 17.5 \\
$M_z$ & 0.44 & 0.11 & 0.71 & 0.15 \\
$N_g$ & 29.6 & 7.4 & 29.4 & 8.4 \\
$N_r$ & 29.6 & 7.7 & 29.6 & 8.0 \\
\hline
\end{tabular}
\caption{The means and dispersions in the properties of the RR Lyrae variables 
  split according to RRab and RRc types. Listed are the amplitudes in the $g$ 
  and $r$ bands, the period, the reduced $\chi^2$ in $g$ and $r$, the Stetson 
  index, the metallicity, the heliocentric distance $D$ and Galactocentric 
  distance $r$, the absolute magnitude $M_z$ and the number of good epochs in 
  $g$ and $r$.}
\label{table:lyr_props}
\end{table}

\section{Substructure revealed by the RR Lyraes}
\label{sect:lyr_subst}

\subsection{RR Lyrae Distances}
\label{sect:rrlyrae_absmagdist}

RR Lyraes are ``standard candles'' because they have a well-defined absolute 
magnitude, which, nonetheless, depends on metallicity.  We calculate distances 
via the distance modulus:
\begin{equation}
\log D = \frac{m_z-M_z+5}{5}
\end{equation}
The apparent magnitudes $m_z$ come directly from the HLC.  The absolute 
magnitudes $M_z$ are obtained via the following relation from \cite{Ca08}:
\begin{eqnarray}
  M_z &=& 1.3706 + 0.8941\log Z + 0.1315 [\log Z]^2 \nonumber \\
  &-& (2.6907 + 0.8192 \log Z + 0.0664 [\log Z]^2) {\rm ln} C_0 \\
  &+& (47.9836 + 31.7879 \log Z + 5.2221 [\log Z]^2) [{\rm ln} C_0]^2
  \nonumber \\ &+& (141.7704 + 100.6676 \log Z + 17.4277 [\log Z]^2)
  [{\rm ln} C_0]^3 \nonumber \\ &+& (0.3286 + 2.0377 \log Z + 0.3882
  [\log Z]^2) \log P \nonumber
\end{eqnarray}
where $P$ is the fundamental period, $\log Z = {\rm [Fe/H]} -1.5515$
and $C_0 = (u\!-\!g)_0 - (g\!-\!r)_0$, with the 0 subscript denoting
that the colours are unreddened. The intrinsic scatter in this
relation is small compared to the errors. Uncertainties on the
metallicities [Fe/H] and distances $D$ are computed using standard
methods.  From this, we find that the distance errors are typically
around $8$ per cent (this includes the scatter due to the metallicity
relationships).

The right ascension, declination, classification, mean magnitudes, amplitude, 
period and distance of all our 407 RR Lyrae candidates are given in an 
accompanying electronic table.  The means and dispersions for some useful 
quantities for the RRab and RRc subsamples are given in 
Table~\ref{table:lyr_props}.

\subsection{The Sagittarius Stream, the Hercules-Aquila Cloud and the
  Pisces Overdensity}

\begin{figure}
\epsfig{file=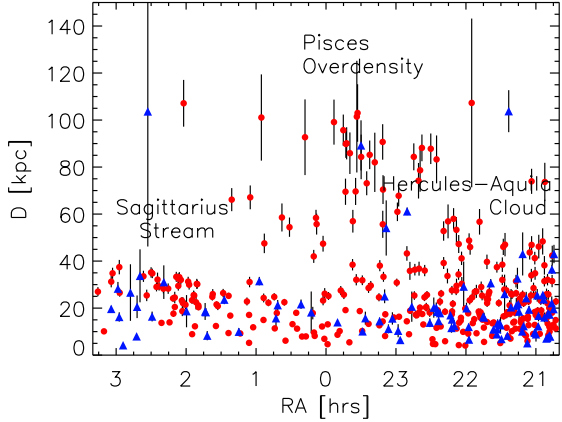,angle=0,width=\linewidth}
\caption{The spatial distribution of the ab-type (red circles) and the
  c-type (blue triangles) RR Lyraes.}
\label{fig:lyr_radist}
\end{figure}

The distribution of RR Lyraes in right ascension and distance is shown in 
Figure~\ref{fig:lyr_radist}, with the ab-types plotted as red circles and the 
c-types plotted as blue triangles. There are a number of things to notice. 
First, there are 296 RR Lyraes at right ascensions $20.7^{\mbox{\small h}} < 
\alpha < 24^{\mbox{\small h}}$, but only 111 at $0^{\mbox{\small h}} <\alpha 
<3.3^{\mbox{\small h}}$. The greatest concentration of RR Lyraes is in the 
fields coincident with the Hercules-Aquila Cloud~\citep{Be07}. Of course, not 
all these RR Lyraes are necessarily associated with the Cloud, as there may be 
contamination from the underlying smooth population associated with the 
Galactic Spheroid. It is known that the Bulge and Spheroid harbour a 
population of RR Lyraes, distributed in a roughly spherical manner around the 
Galactic Centre, with a metallicity distribution peaked at [Fe/H] $\sim -1$
~\citep[see e.g.][]{Wa91,Al98,Co06}.  The plane of the orbit of the 
Sagittarius dwarf galaxy crosses Stripe 82, and there is a visible overdensity 
of RR Lyraes at this location ($\alpha \approx 2^h$). Finally, we note that 
there are few (specifically 47) RR Lyrae at large distances ($D$$>$ 50\,kpc), 
of which 28 lie in a clump at $\alpha \approx 23.5^{\mbox{\small h}}$.We term 
the structure the {\it Pisces Overdensity}. Distance uncertainties are shown 
as vertical bars for each RR Lyrae -- although the error bars for the distant 
RR Lyrae are large enough to be visible, they cannot be responsible for the 
overdensity.
\begin{figure}
\begin{center}
\epsfig{file=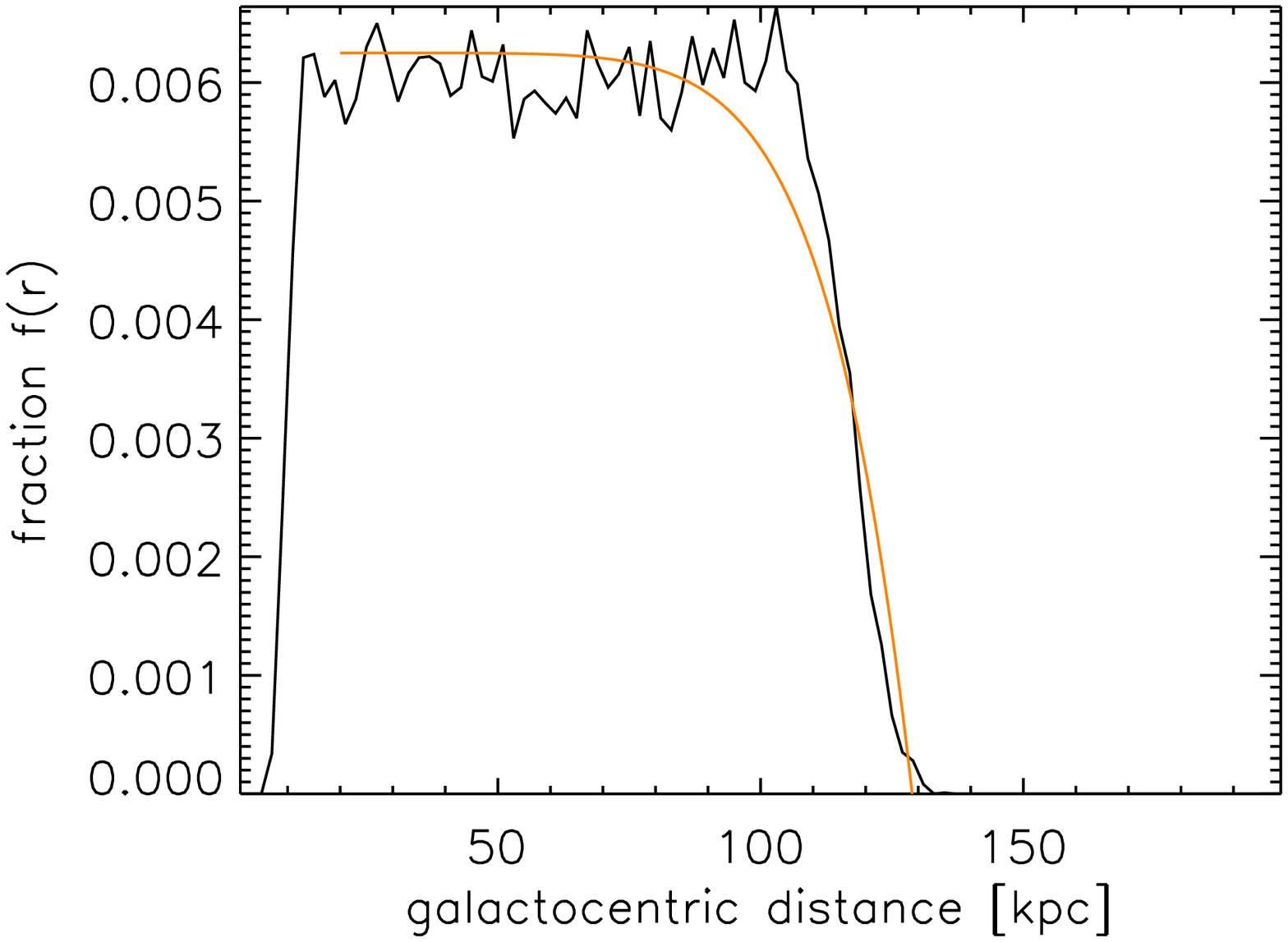,angle=0,width=0.8\linewidth}
\epsfig{file=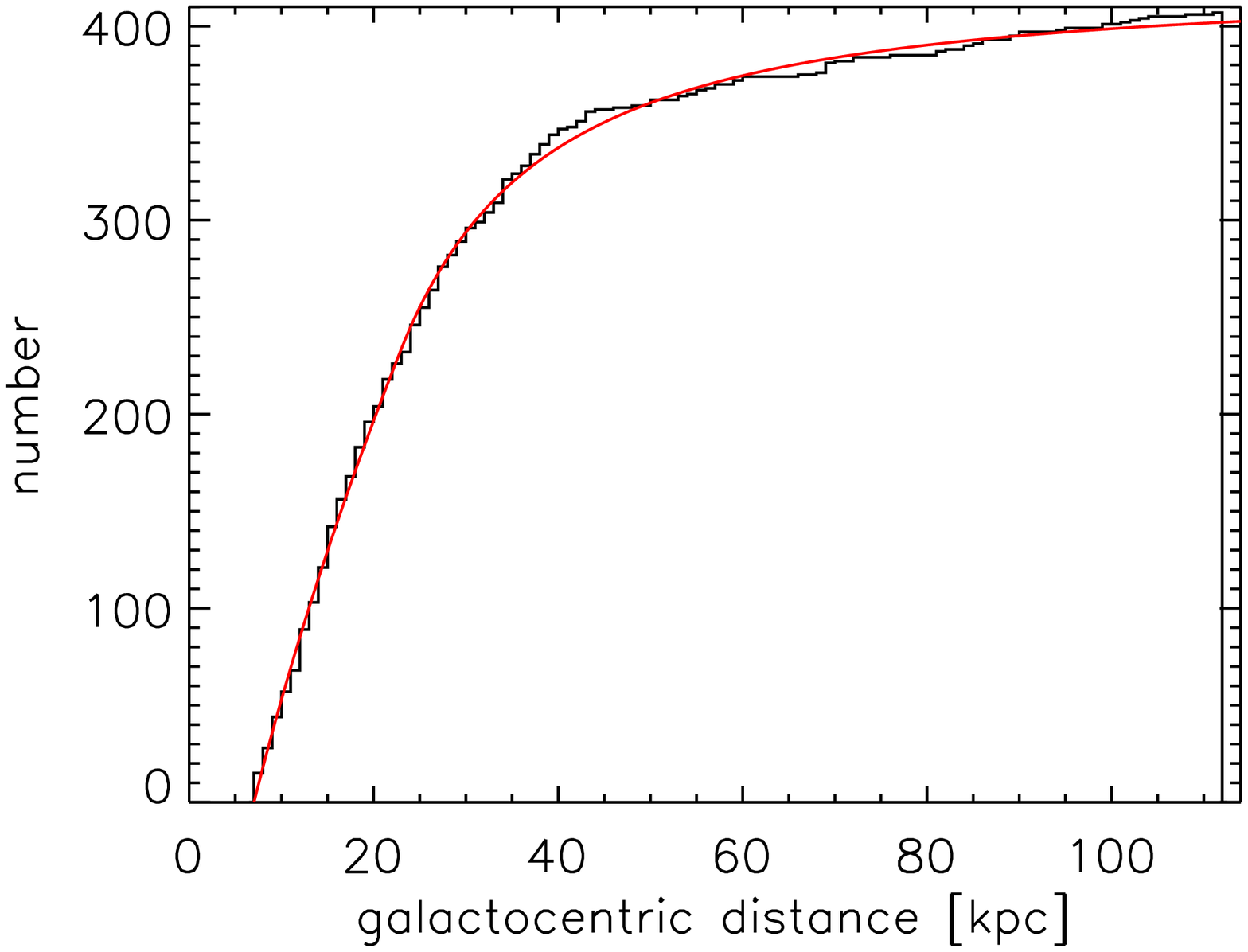,angle=0,width=0.8\linewidth}
\caption{Top: The fraction of Galactic volume probed by the RR Lyraes
  in Stripe 82 as function of Galactocentric distance $r$. Bottom: The
  cumulative number of RR Lyrae within radius $r$ in Stripe 82,
  together with the fits given in eqn~(\ref{eq:denslaw}).}
\label{fig:denslaw}
\end{center}
\end{figure}
\begin{figure}
\epsfig{file=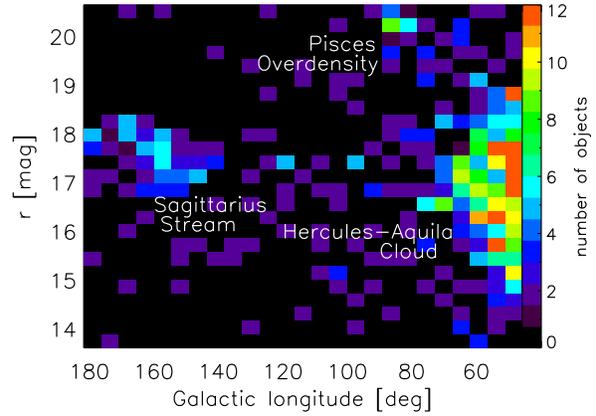,angle=0,width=\linewidth}
\caption{The number density of RR Lyraes in the plane of Galactic
  longitude versus $r$ magnitude. There are three obvious
  overdensities corresponding to the Sagittarius Stream, the
  Hercules-Aquila Cloud and the Pisces Overdensity.}
\label{fig:prominent}
\end{figure}
\begin{table*}
\centering
\begin{tabular}{lcccccc}
  \hline
  Substructure & $\ell$ (deg) & $b$ (deg) & $D$ (kpc) & $r$ (kpc) &
  $P$ (days) & $[{\rm Fe/H]}$ \\
  \hline
Hercules-Aquila Cloud & [45$^\circ$,79$^\circ$] & [-56$^\circ$,-24$^\circ$] & 21.9$\pm$12.1 & 20.1$\pm$11.3 & 0.51$\pm$0.12 & -1.43$\pm$0.36 \\
Sagittarius Stream & [139$^\circ$,182$^\circ$] & [-62$^\circ$,-46$^\circ$] & 26.1$\pm$ 5.6 & 31.4$\pm$ 5.6 & 0.54$\pm$0.12 & -1.43$\pm$0.30 \\
Pisces Overdensity & [63$^\circ$,93$^\circ$] & [-60$^\circ$,-46$^\circ$] & 79.9$\pm$13.9 & 79.4$\pm$14.1 & 0.56$\pm$0.08 & -1.48$\pm$0.28 \\
  \hline
\end{tabular}
\caption{The range in Galactic coordinates and the means and dispersions in 
  heliocentric distance, Galactocentric distance, period and metallicity of 
  the RR Lyrae variables in the Hercules-Aquila Cloud, the Sagittarius Stream 
  and the Pisces Overdensity.}
\label{table:lumpprops}
\end{table*}
\begin{figure}
\epsfig{file=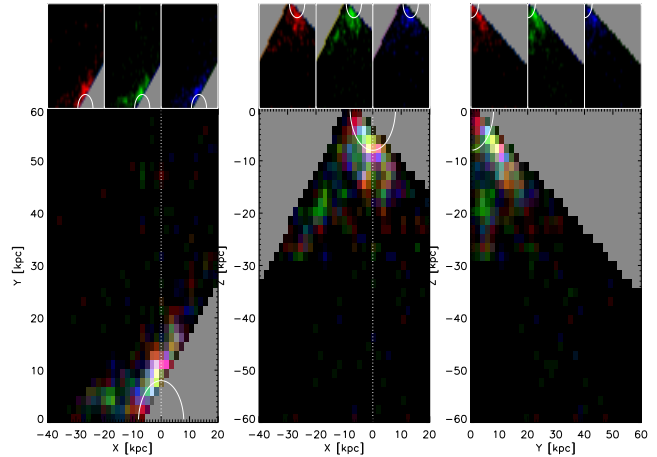,angle=0,width=\linewidth}
\caption{The density distributions of RR Lyrae candidates projected onto the 
  principal planes of Galactocentric $(x,y,z$) coordinates. Here, red 
  represents the high metallicity RR Lyrae population ([Fe/H] $> -1.3$), green 
  medium ($-1.5 <$ [Fe/H] $< -1.3$) and blue low ([Fe/H] $< -1.5$). The insets 
  show each distribution plotted separately, together with a white circle 
  centered at the origin. There is no obvious distinction according to 
  metallicity, and the low, medium and high metallicity RR Lyraes are clearly 
  distended and seemingly belong to the Hercules-Aquila Cloud. In particular, 
  none of the populations is distributed in a spherically symmetric manner 
  around the Galactic Centre.}
\label{fig:threed}
\end{figure}

The upper panel of Figure~\ref{fig:denslaw} shows the fraction of accessible 
volume as a function of Galactocentric radius $r$ probed by our survey. The 
volumes are calculated via Monte Carlo integrations in which the RR Lyrae 
luminosity function is modelled as a Gaussian with mean and dispersion from 
Table~\ref{table:lyr_props}; the magnitude limits used were those defining our 
RR Lyrae sample. The survey reaches at least $r$$\sim$100\,kpc before the 
accessible volume begins to decline. The brightest RR Lyraes in our sample 
have $M_z$=0.1 and so are still detectable within $r$$\sim$130\,kpc. In 
classical models of the smooth halo, the RR Lyraes are distributed as a power-
law like $\rho \sim r^{-n}$ with $n$$\sim$3.1 (Wetterer \& McGraw 1996).  With 
no substructure present in Stripe 82, the right ascension-distance graph of 
Figure~\ref{fig:lyr_radist} would look rather different. The fall-off in 
numbers would be steady, and not as sharp as the drop observed beyond 
$D$$\sim$40\,kpc, which is real and cannot be attributed to properties of the 
survey.

\begin{figure*}
\epsfig{file=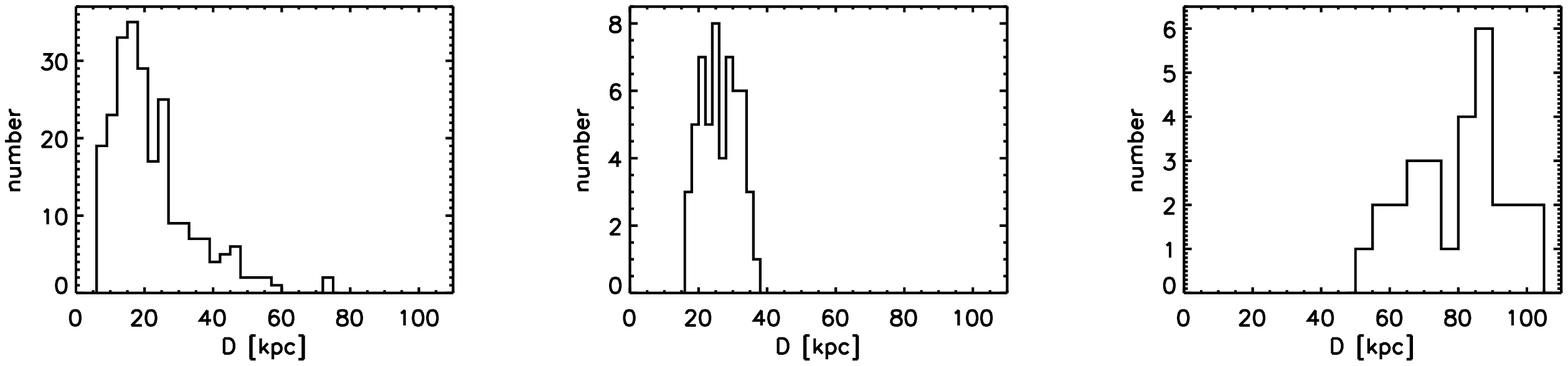,angle=0,width=0.8\linewidth}
\epsfig{file=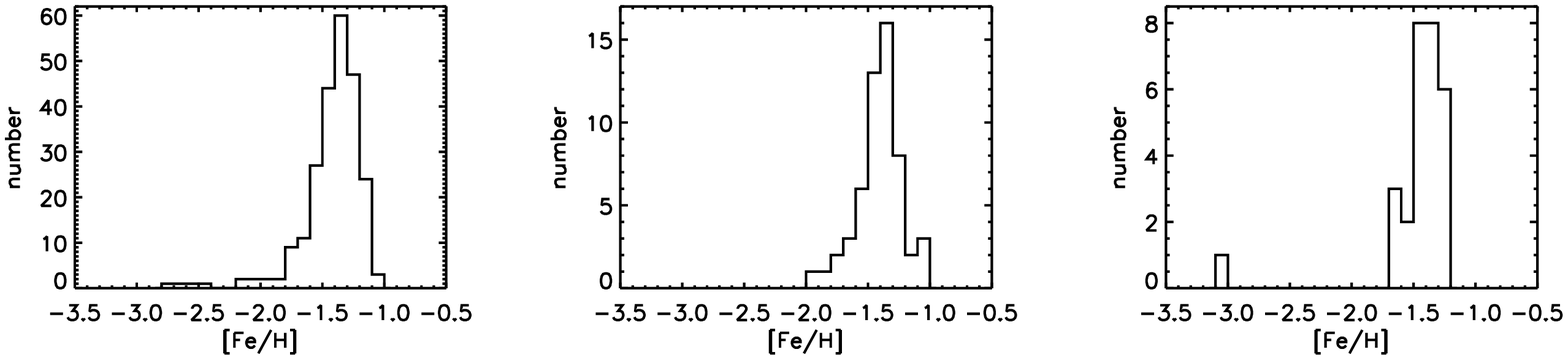,angle=0,width=0.8\linewidth}
\caption{The distributions of distances (top) and metallicities
  (bottom) of RR Lyraes in the Hercules-Aquila Cloud (left)
  Sagittarius Stream (centre) and the Pisces Overdensity (right).}
\label{fig:disdis}
\end{figure*}

The presence of an edge to the RR Lyrae distribution in the stellar halo at 
$r\sim$50\,kpc has been proposed before by~\citet{Iv00}, using a sample of 148 
RR Lyraes in SDSS commissioning data. However, the same authors later applied 
their method to a larger area of the sky and found no break until at least 
70\,kpc~\citep{Iv04}. Vivas et al. (2006) also found no break before the limit 
of their survey at $\sim$60\,kpc. So, ``edge'' may be too strong a term, but 
the number density profile of the RR Lyraes does seem to be best matched by a 
broken power-law, as shown in the lower panel of Figure~\ref{fig:denslaw}. 
Such a parameterisation was first advocated by Saha (1985), who noticed that 
the RR Lyrae density fell off much more rapidly beyond Galactocentric radii of 
25\,kpc.  Adjusting by the fraction of Galactic volume sampled by our survey, 
and assuming that our efficiency is $\epsilon \approx 1$, we find that the 
spherically-averaged number density of RR Lyrae as
\begin{equation}
    n(r) = 2.6 \times 
\begin{cases} \quad \left( {\displaystyle 23\,
        {\rm kpc}  \over \displaystyle r} \right)^{2.4} & \text{ if
      $5 < r
      \le 23$ kpc}  \\ \quad
 \left( {\displaystyle 23\, {\rm kpc} \over \displaystyle r} \right)^{4.5} & \text{ if $23 < r < 100$ kpc}
\end{cases}
\label{eq:denslaw}  
\end{equation}
out to $\sim$100\,kpc, beyond which our data is highly incomplete, with only 
bright RR Lyraes detectable (see Figure~\ref{fig:disdis}). Our break radius of 
23\,kpc is very close to that proposed by Saha (1985).

The inner power-law slope is almost the same as that found by Miceli et al 
(2008) -- namely $n$=-2.43 -- in a very large sample of RR Lyrae stars closer 
than 30\,kpc in the LONEOS survey. Of course, formulae such as 
eqn~(\ref{eq:denslaw}) are just a parameterisation of the data, as the RR 
Lyrae distribution is neither spherically symmetric nor smooth, but dominated 
by the three structures in the Stripe. The break at $r$$\sim$25\,kpc is really 
a consequence of the fact that most of the RR Lyraes are in the Hercules-
Aquila Cloud and the Sagittarius Stream substructures, which lie within 
40\,kpc of the Galactic centre.  A similar conclusion regarding the importance 
of substructure is reached by \citet{Se07}, who divide their RR Lyrae 
distribution into 13 clumps, of which they suggest at least seven correspond 
to real substructures.

In Figure~\ref{fig:prominent}, the number density of RR Lyraes is plotted in 
the plane of Galactic longitude versus $i$ band magnitude. The three 
substructures show up very clearly, together with some isolated hot pixels 
that may be indicators of real objects. We define the Sagittarius Stream RR 
Lyraes via
\begin{equation}
180^\circ >\ell > 135^\circ, \qquad 16.5 < r < 18.5
\end{equation}
RR Lyraes associated with the Hercules-Aquila Cloud are extracted via
\begin{equation}
80^\circ > \ell > 45^\circ, \qquad 14.5 < r < 20, \qquad \ell + 15 r < 358
\end{equation}
For the Pisces Overdensity, we chose the stars satisfying
\begin{equation}
95^\circ > \ell > 60^\circ, \qquad r > 19, \qquad \ell + 15 r > 358.
\end{equation}
These cuts give 55 stars in the Sagittarius Stream, 28 in the Pisces
Overdensity, and 237 in the Hercules-Aquila Cloud.

Ideally, we would like to separate any contaminating Bulge and
Spheroid RR Lyrae from those of the Hercules-Aquila Cloud, but this is
not easy. In particular, Figure~\ref{fig:threed} shows the density
distribution of the RR Lyrae populations colour-coded according to
metallicity. The comparatively metal-rich RR Lyraes (red in the
Figure) do not seem to be distributed any differently from the
comparatively metal-poor (green and blue).  In fact, all the
distributions are distended and distributed asymmetrically relative to
the Galactic Centre, consistent with the bulk of the stars belonging
to the Hercules-Aquila Cloud.

The properties of the RR Lyrae in the different substructures are
listed in Table~\ref{table:lumpprops}. Note that very nearly 60\,per
cent of all the RR Lyraes in Stripe 82 are associated with the
Hercules-Aquila Cloud, emphasising the arguments made by \citet{Be07}
as to the importance of this structure. The mean heliocentric
distances of the Hercules-Aquila Cloud and the Sagittarius Stream are
comparable in Stripe 82, but the Pisces Overdensity is much further
away at $D$$\sim$80\,kpc. The Pisces Overdensity lies within a few
degrees of the Magellanic Plane. Although the distance of the
Overdensity is greater than that of the Large and Small Magellanic
Clouds ($D$$\sim$55\,kpc), it is possible that the Magellanic Stream
may be more distant in this area of the sky. Thus, at present, it is
unclear whether the Pisces Overdensity is related to Magellanic Cloud
debris.

\begin{figure}
\epsfig{file=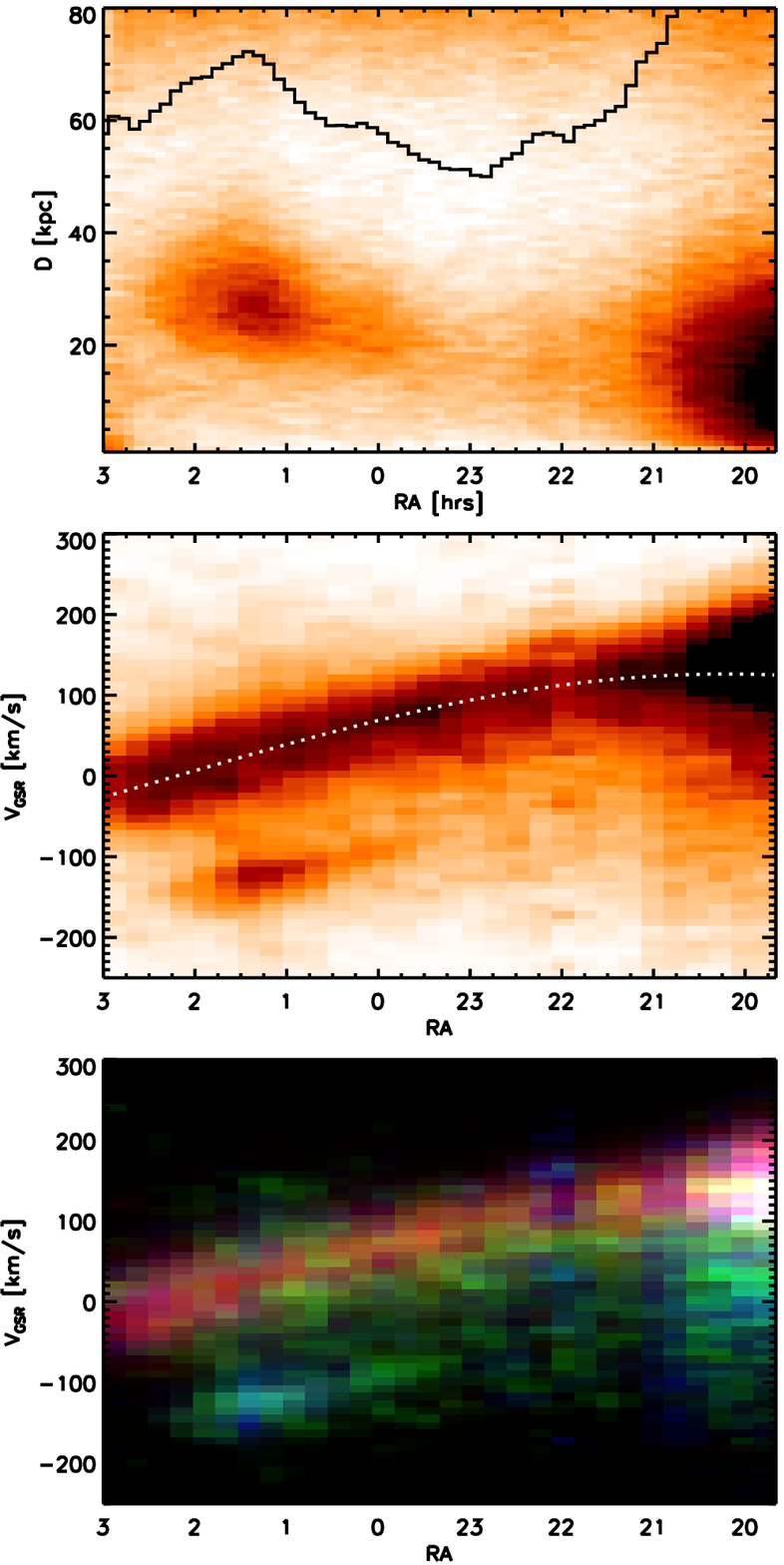,angle=0,width=0.9\linewidth}
\caption{Top: The distributions of main sequence turn-off stars in
    Stripe 82 shown in the plane of right ascension versus
    heliocentric distance. The black histogram shows the
    one-dimensional distribution as a function of right ascension
    only. Middle: The velocity distribution of all stars with $g-i<1$
    with SDSS spectra. The white dashed curve shows the line $190
    \sin \ell \cos b$ kms$^{-1}$ and marks the expected locus of stars
    belonging to the thick disk. Most thin disk stars are excised by
    the colour cut. Bottom: As middle, but the distribution is now
    colour-coded according to metallicity with red representing [Fe/H]
    $> -1$, green $-1.67 <$ [Fe/H] $< -1$ and blue low [Fe/H] $<
    -1.67$. This separates the thin disk stars (reddish) from the
    older and metal-poor components of the stellar halo, such as the
    Hercules-Aquila Cloud.}
\label{fig:vasily}
\end{figure}
\begin{figure}
\epsfig{file=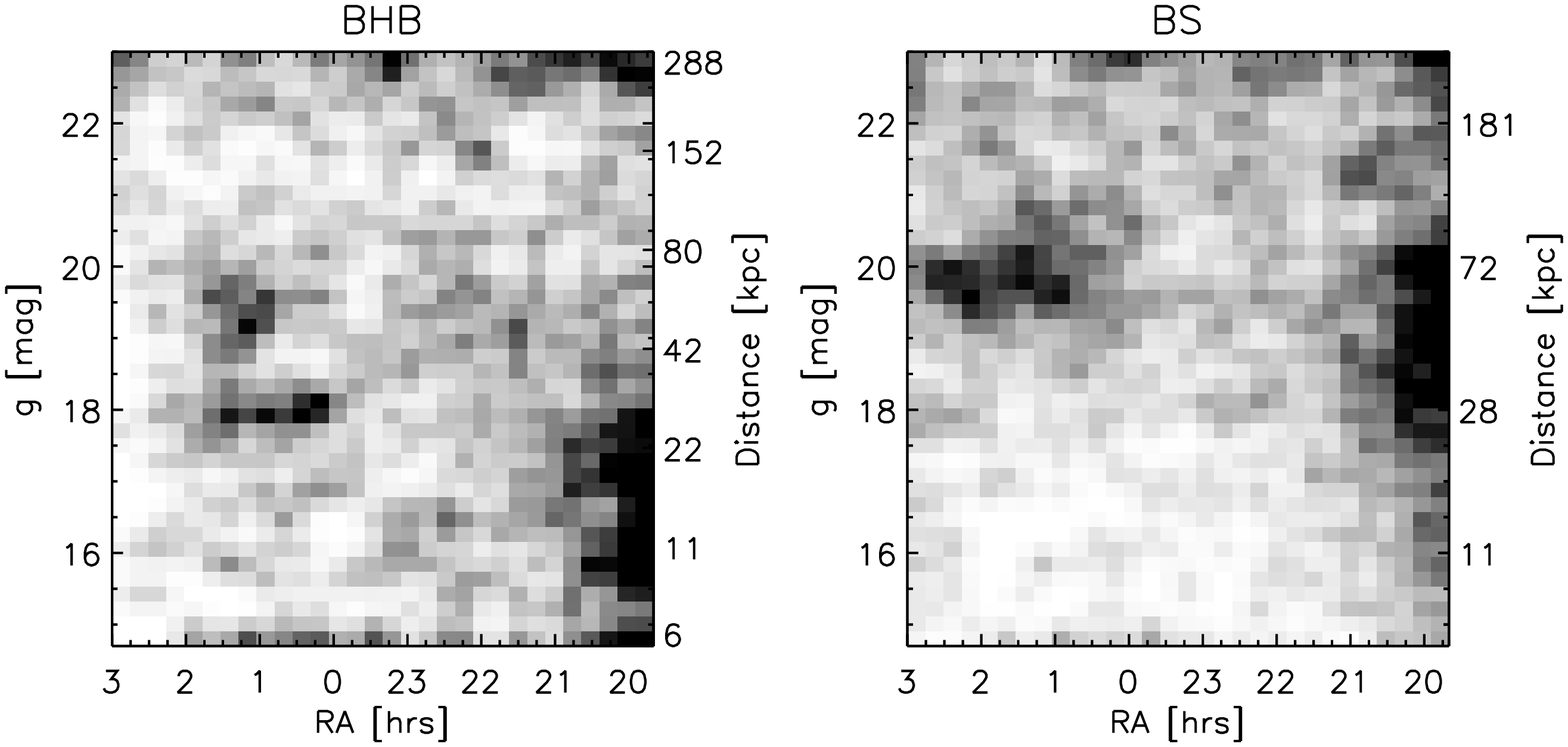,angle=0,width=\linewidth}
\caption{The distributions of BHB stars (left) and blue stragglers
  (right) in Stripe 82 shown in the plane of right ascension versus
  heliocentric distance.}
\label{fig:vasilyb}
\end{figure}

One way to estimate the mass of the Pisces Overdensity is to compare with the 
Carina dwarf spheroidal, which is at a similar distance \citep[$\sim$100
\,kpc; ][]{Ma98}. Carina has a total mass of $\sim 2 \times 10^7 M_\odot$ and 
a mass-to-light ratio of $\sim$70. At least 75 RR Lyrae stars have been 
detected by~\citet{Da03} using well-sampled multi-epoch data in the $B$ and 
$V$ bands, although over a small baseline of a few days.  Assuming similar 
stellar populations and similar efficiencies of detection of bright RR Lyrae 
in the surveys, then we can use a simple scaling argument to estimate the 
total stellar mass associated with the Pisces Overdensity as $\sim 10^5 
M_\odot$.  We can corroborate this mass estimate by comparison with our data 
on the Hercules-Aquila Cloud. The calculation using the Hercules-Aquila Cloud 
has the advantage that the RR Lyrae populations in the two structures have 
been discovered by the same algorithm, but the disadvantage that the 
properties of the Cloud are also rather uncertain. The absolute magnitude of 
the Cloud is given by~\citet{Be07} as $M_r$=-13, suggesting that its total 
stellar mass is $\sim 10^7 M_\odot$. Of course, the Cloud is an enormous 
structure, covering $\sim$80$^\circ$ in longitude and probably extending above 
and below the Galactic plane by 50$^\circ$.  Only a small fraction ($\sim$
1\,per cent) of the Cloud is probed by the Stripe 82 data, suggesting that 
there must be $\sim 2 \times 10^4$ RR Lyraes associated with the Cloud in 
total. The mass of the Cloud covered by the Stripe is $\sim 10^5 M_\odot$. 
Again, assuming similar stellar populations, the mass associated with the 
Pisces Overdensity is at least $\sim 10^4 M_\odot$, a value encouragingly 
similar to our first estimate.

Of the 28 stars identified as members of the Pisces Overdensity, the 
intrinsically faintest has $M_z = 0.76$. The stars extend over an area 
of 55\,deg$^2$ of Stripe 82. Thus, the surface number density of the RR
Lyrae is 0.51\,deg$^{-2}$. By comparison, the Hercules-Aquila Cloud 
has 209 RR Lyraes with an absolute magnitude brighter than $M_z$=0.76. 
They cover 95\,deg$^2$ of Stripe 82, and hence the surface number density 
is 2.20\,deg$^2$. This suggests that the Pisces Overdensity is 
$\sim$4.33 times more diffuse than the Hercules-Aquila Cloud.

The distance and metallicity distributions of our RR Lyraes can be
used to study the properties of the Hercules-Aquila Cloud, the
Sagittarius Stream and the Pisces Overdensity. Plots are shown in
Figure~\ref{fig:disdis}. The distribution of heliocentric distances
for the Cloud has a mean of $22.0$\,kpc and a standard deviation of
$12.1$\,kpc. One possible interpretation of the Cloud is that it is
analogous to caustic features like the shells seen around elliptical
galaxies. However, the considerable depth of the Cloud seen in the RR Lyrae
distribution tends to argue against such an interpretation as a
caustic structure.

In directions towards Stripe 82, the distances of the arms of the Sagittarius 
Stream are not well-known. Simulations offer a rough guide, but no more than 
that. The upper panel of Figure 3 of \citet{Fe06} shows the young leading arm 
(A), together with parts of the old trailing arm (B), at distances of 15-20
\,kpc; further, material belonging to parts of B, the old leading arm (C) and 
the young trailing arm (D) spread out over a swathe of distances 30-60\,kpc.  
The distribution of distances of our Sagittarius RR Lyraes in Figure
~\ref{fig:disdis} does indeed show some evidence of bimodality. It is possible 
that the peak at distances $D$$\sim$20\,kpc corresponds to the A and B 
streams, whilst the peak at $D$$\sim$35\,kpc corresponds to the other wraps.  
However, the simulations also suggest that the second peak should be much 
broader than appears to be the case in the data.  Our identifications are 
tentative and radial velocity data is required to enable a cleaner separation 
of the wraps, as is evident from the lower panel of Figure 3 of Fellhauer et 
al. (2006). 

From Table~\ref{table:lumpprops}, we see that the RR Lyrae in the Hercules-
Aquila Cloud have a metallicity [Fe/H]=-1.42$\pm 0.24$, whilst those 
associated with the Sagittarius Stream have [Fe/H]= -1.41$\pm 0.19$.  The 
Pisces Overdensity has a metallicity [Fe/H]= -1.47$\pm 0.34$, comparable to 
the Hercules-Aquila Cloud, but richer than the typical populations in the 
outer halo, which have a metallicity of [Fe/H]$\sim$-2~\citep{CB07}. 

In fact, \citet{Vi05} have already carried out VLT spectroscopy of 14
RR Lyrae variables that lie in the leading arm of the Sagittarius
Stream, finding a metallicity of [Fe/H]=-1.76$\pm 0.22$. The stars
lie well away from Stripe 82 at right ascensions $13^{\mbox{\small h}}
< \alpha < 16^{\mbox {\small h}}$ and at heliocentric distances of
$\sim$50\,kpc. Many of the RR Lyraes in our sample will belong to the
trailing arm, which may account for some of the difference in the
metallicity estimate.

We show a view of Stripe 82 as derived from SDSS main-sequence
turn-off (MSTO) stars in Figure~\ref{fig:vasily}.  The upper panel
gives the number of MSTO stars as a function of right ascension and
distance. The one-dimensional histogram plotted in black shows the
dependence of number on right ascension alone. Note that the
Sagittarius Stream is immediately visible at $\alpha \approx
40^\circ$. There are two density maxima in the black histogram,
perhaps hinting that more than one wrap of the Stream is detectable in
MSTO stars. The Hercules-Aquila Cloud substructure also shows up very
clearly, although the fainter Pisces Overdensity is understandably
absent. The distance estimates to the substructures derived
from MSTO stars agree well with those from RR Lyraes. The SDSS DR6
includes a large number of stellar spectra which have been analysed to 
provide radial velocities and fundamental stellar atmospheric parameters
\citep{Le08}. The velocities of all stars with spectra and satisfying
$ g-i<1$, to remove most of the thin disk contaminants, are plotted
against right ascension along Stripe 82 in the middle panel. Of
course, most of the stellar targets are disk stars, and so the curve
$v_{\rm GSR} = 190 \cos b \sin \ell$ kms$^{-1}$ is plotted to show the
locus of the thick disk in this dataset.  The Sagittarius
Stream stars are clearly offset in velocity from the thin disk at
$v_{\rm GSR} \approx -130$ kms$^{-1}$. The bottom panel shows the same
data, but now colour-coded according to metallicity so as to highlight
different structures. We can detect the kinematically bifurcated
Sagittarius Stream and clearly see the separation of the more
metal-rich Galactic disk and bulge stars from the Hercules-Aquila
Cloud. The eye can also discern some fainter substructure, the reality of
which remains to be established.

A final view of Stripe 82 is provided in Figure~\ref{fig:vasilyb},
which shows the density distribution of blue horizontal branch stars
(BHBs) and blue stragglers (BSs), selected using the colour cuts of
\cite{Ya00}. The BHB population of course abut the RR Lyrae population
in the Hertzsprung-Russel diagram. We might expect to see all three
substructures in the BHB density -- and so it is reassuring that the
Sagittarius stream, the Hercules-Aquila cloud and the Pisces
Overdensity are all visible. The same substructures are also
identifiable in the BS populations with the exception of the Pisces
Overdensity which is of course too distant. The Sagittarius
Stream is clearly bifurcated in BHBs, although not in BSs, suggesting
that the ratio of BHBs to BSs varies along the Stream. There is some
evidence for bimodality in the BHB distance distribution of the
Hercules-Aquila Cloud.

\section{Conclusions}
\label{sect:concl}

We have constructed a catalogue of 21\,939 variable objects in Stripe 82.
The catalogue of variables is published in full as an electronic
supplement to this article.  We have extracted a sample of RR Lyrae
stars, 316 RRab types and RRc types, from the variable catalogue,
using a combination of cuts based on colour, period, amplitude and
metallicity. The RR Lyraes lie at distances 5-115\,kpc from the
Galactic centre and individual distance estimates, accurate to
typically 8 per cent, are calculated using the colour, period and
metallicity to estimate absolute magnitude.

If the RR Lyrae data are modelled by a smooth density distribution,
then a good fit is provided by a broken power-law. The number density
of RR Lyrae falls with Galactocentric radius $r$ like $n(r) \sim
r^{-2.4}$ for $5 < r < 23$ kiloparsecs, switching to a much steeper
decline, $n(r) \sim r^{-4.5}$ for $23 < r < 100$ kiloparsecs.
However, smooth, spherically-averaged density laws do not tell the
whole story, as in reality the RR Lyrae distribution is strongly
clumped. In Stripe 82, the distribution of RR Lyraes is dominated by
three enormous substructures -- namely, the Hercules-Aquila Cloud, the
Sagittarius Stream and the Pisces Overdensity. We identified samples
of 237 RR Lyraes in the Hercules-Aquila Cloud, 55 stars in the
Sagittarius Stream and 28 in the Pisces Overdensity.

RR Lyraes belonging to the Hercules-Aquila Cloud are very numerous,
and comprise almost 60 per cent of our entire Stripe 82
sample. Although there may be some contamination from a smooth
component of RR Lyraes associated with the Galactic Bulge and
Spheroid, there is no doubt concerning the existence of the structure,
supporting the initial identification of \citet{Be07}. We estimate
that the total number of RR Lyraes associated with the Cloud is $2
\times 10^4$. The Hercules-Aquila RR Lyraes lie at distances from the
Galactic Centre of 20.2$\pm 11.3$\,kpc, and are metal-poor with [Fe/H]
$= -1.42 \pm 0.24$.

Both leading and trailing arms of the Sagittarius Stream also
intersect Stripe 82. Simulations predict that the leading wrap is
closer in heliocentric distance than the trailing, but the locations
of the arms are not accurately known in this region of the sky. The
heliocentric distances of our Sagittarius RR Lyraes, which
predominantly are associated with the trailing arm, have a mean of
26.2\,kpc and a dispersion of 5.5\,kpc, whilst their metallicity is
[Fe/H] = $-1.41 \pm 0.19$.

We have also identified a new concentration -- the {\it Pisces
  Overdensity} -- consisting of 28 RR Lyraes centered on Galactic
coordinates of $(\ell \approx 80^\circ, b \approx -55^\circ$). This is
one of the most distant clumps so far found in the halo, as the RR
Lyrae lie at distances of $\sim$80\,kpc. Although the location
is close to the Magellanic Plane, the Pisces Overdensity is much more
distant than the Magellanic Clouds and may well be unrelated to
any known component of the Galaxy. We have made an
order-of-magnitude estimate of the total mass associated with the
Overdensity as at least $\sim 10^4 M_\odot$. The associated RR Lyrae
have a metallicity [Fe/H]=-1.47$\pm 0.34$, comparable to the
Hercules-Aquila Cloud, but richer than the typical populations in the
outer halo.

Our investigation has exploited the advantages of RR Lyrae stars for
identifying remnants and substructure present in the halo of the
Galaxy. Together with earlier SDSS discoveries~\citep{Be07,Ju08}, the
patchy and clumpy nature of the RR Lyrae distribution adds support to
the picture of an outer halo composed of overdensities and voids,
perhaps entirely devoid of any smooth component
\citep[e.g. ][]{Be08}. Further study of the kinematics and
metallicities of RR Lyraes in Stripe 82 should lead to a major advance
in our understanding of the Galactic halo, albeit that significant
observational resources will be required to acquire the necessary
follow-up spectroscopy.

\section*{Acknowledgments}

We thank an anonymous referee for a thoughtful reading of the paper, as well as
Nathan de Lee for generously communicating some of his results in advance of
publication.  NWE, GFG, PCH and DZ acknowledge support from the STFC-funded
Galaxy Formation and Evolution programme at the Institute of Astronomy.  VB
is supported by a Royal Society University research Fellowship.

Funding for the SDSS and SDSS-II has been provided by the Alfred P.
Sloan Foundation, the Participating Institutions, the National Science
Foundation, the U.S. Department of Energy, the National Aeronautics
and Space Administration, the Japanese Monbukagakusho, the Max Planck
Society, and the Higher Education Funding Council for England. The
SDSS Web Site is http://www.sdss.org/.
                                                                               
The SDSS is managed by the Astrophysical Research Consortium for the
Participating Institutions. The Participating Institutions are the
American Museum of Natural History, Astrophysical Institute Potsdam,
University of Basel, Cambridge University, Case Western Reserve
University, University of Chicago, Drexel University, Fermilab, the
Institute for Advanced Study, the Japan Participation Group, Johns
Hopkins University, the Joint Institute for Nuclear Astrophysics, the
Kavli Institute for Particle Astrophysics and Cosmology, the Korean
Scientist Group, the Chinese Academy of Sciences (LAMOST), Los Alamos
National Laboratory, the Max-Planck-Institute for Astronomy (MPIA),
the Max-Planck-Institute for Astrophysics (MPA), New Mexico State
University, Ohio State University, University of Pittsburgh,
University of Portsmouth, Princeton University, the United States
Naval Observatory, and the University of Washington.

\appendix

\section{Comparison with Other Variability Surveys in Stripe 82}
\label{app:compsesar}

\begin{figure*}
\epsfig{file=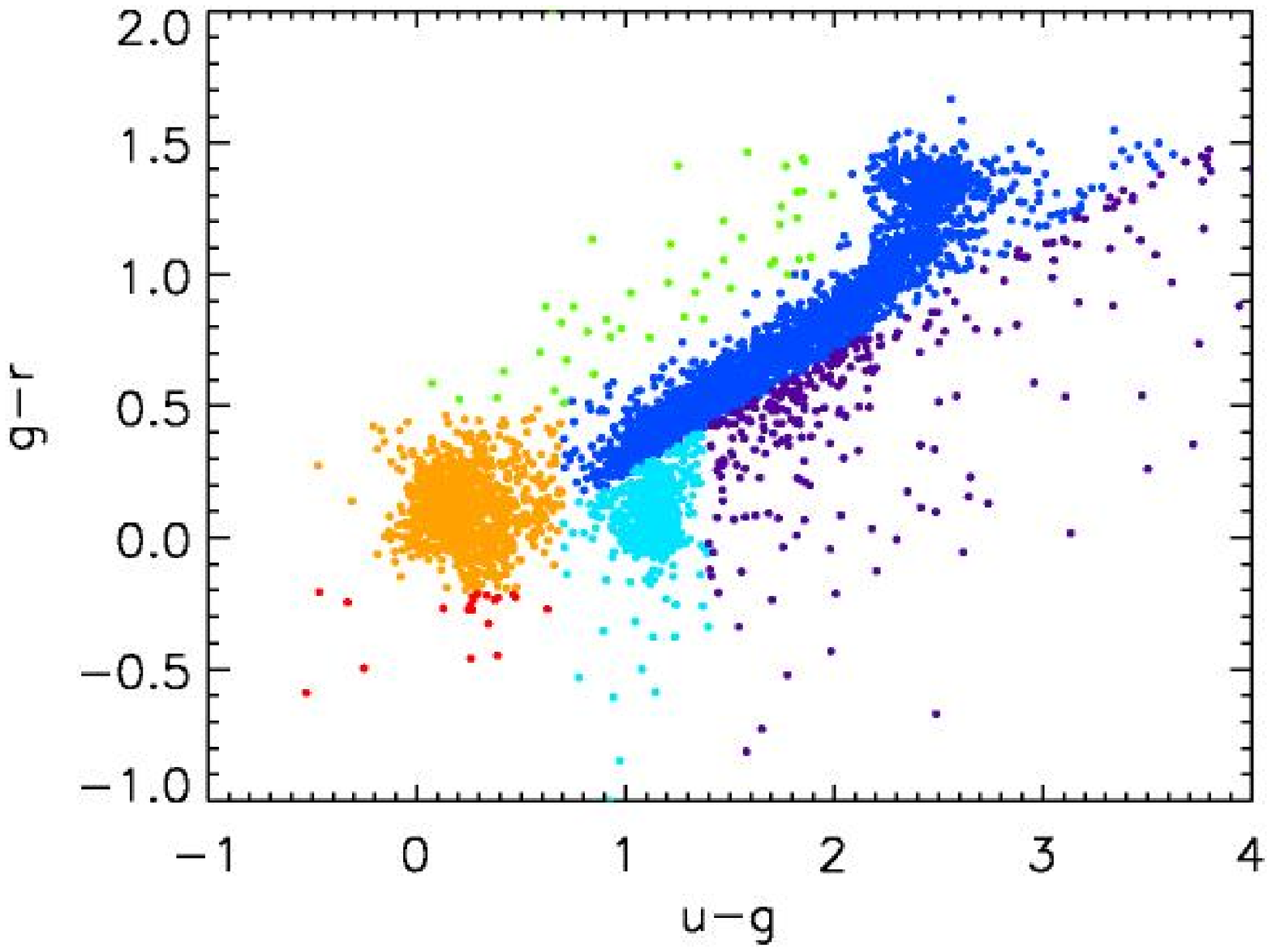,angle=0,width=0.4\linewidth}
\epsfig{file=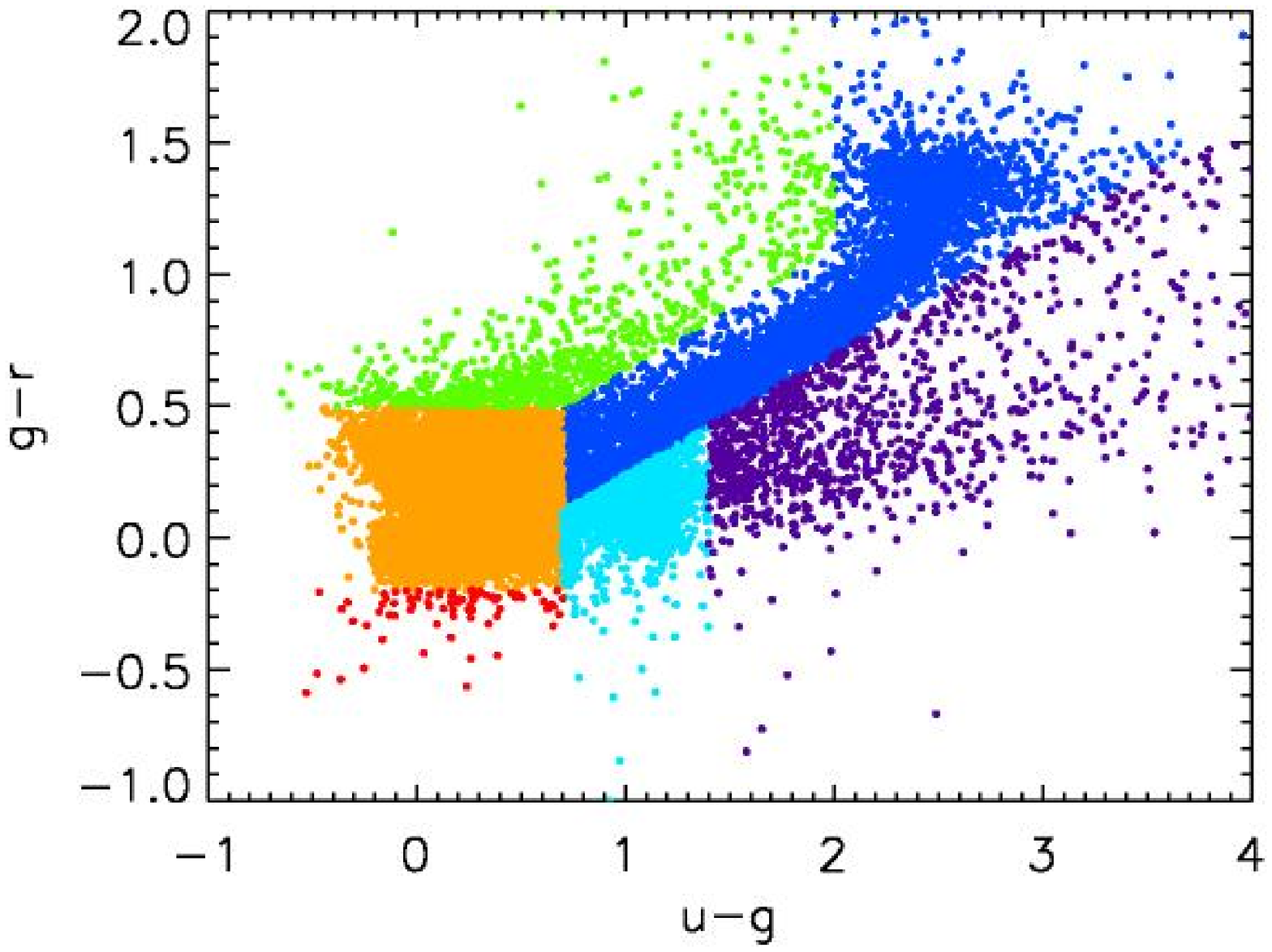,angle=0,width=0.4\linewidth}
\epsfig{file=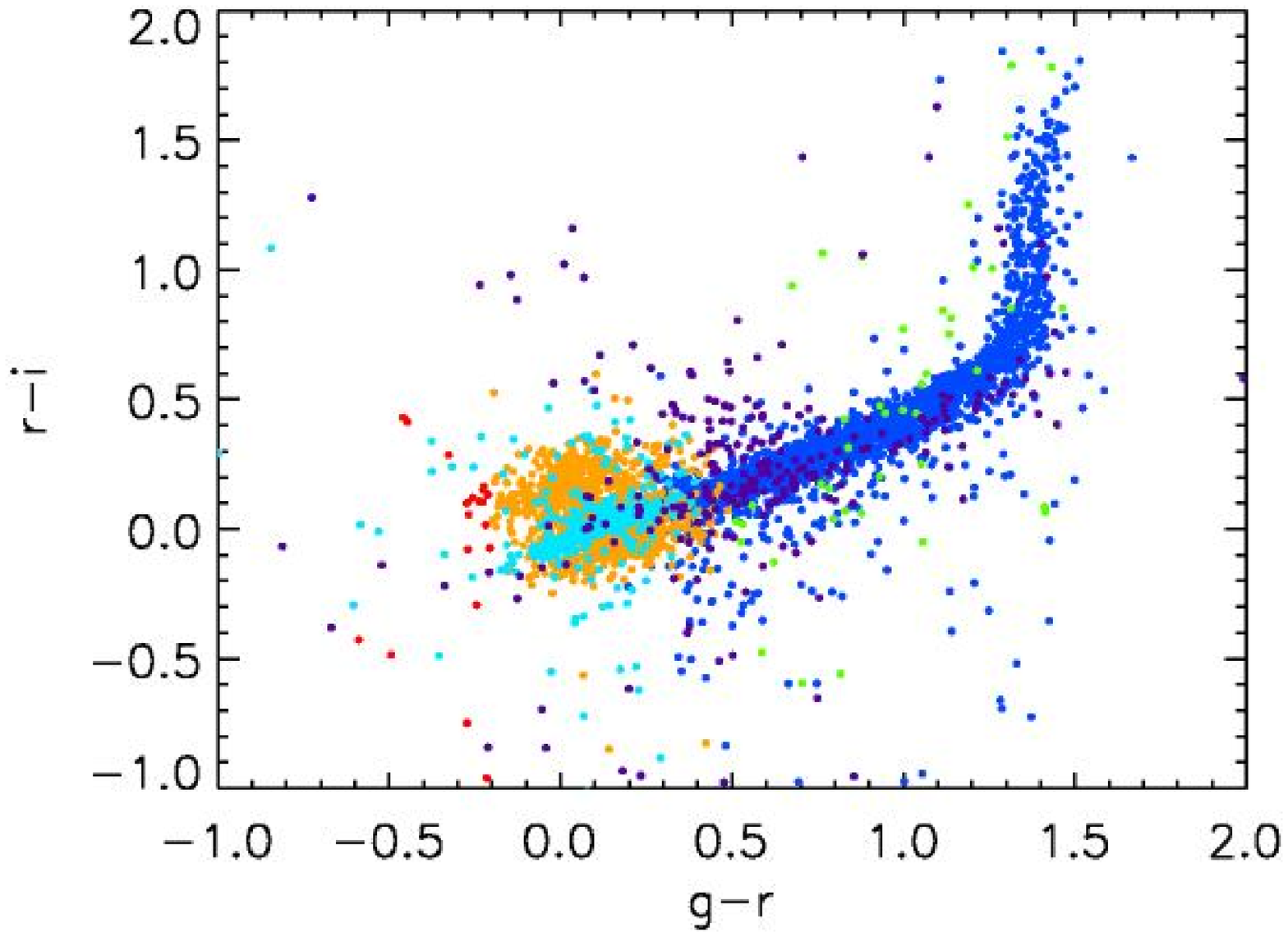,angle=0,width=0.4\linewidth}
\epsfig{file=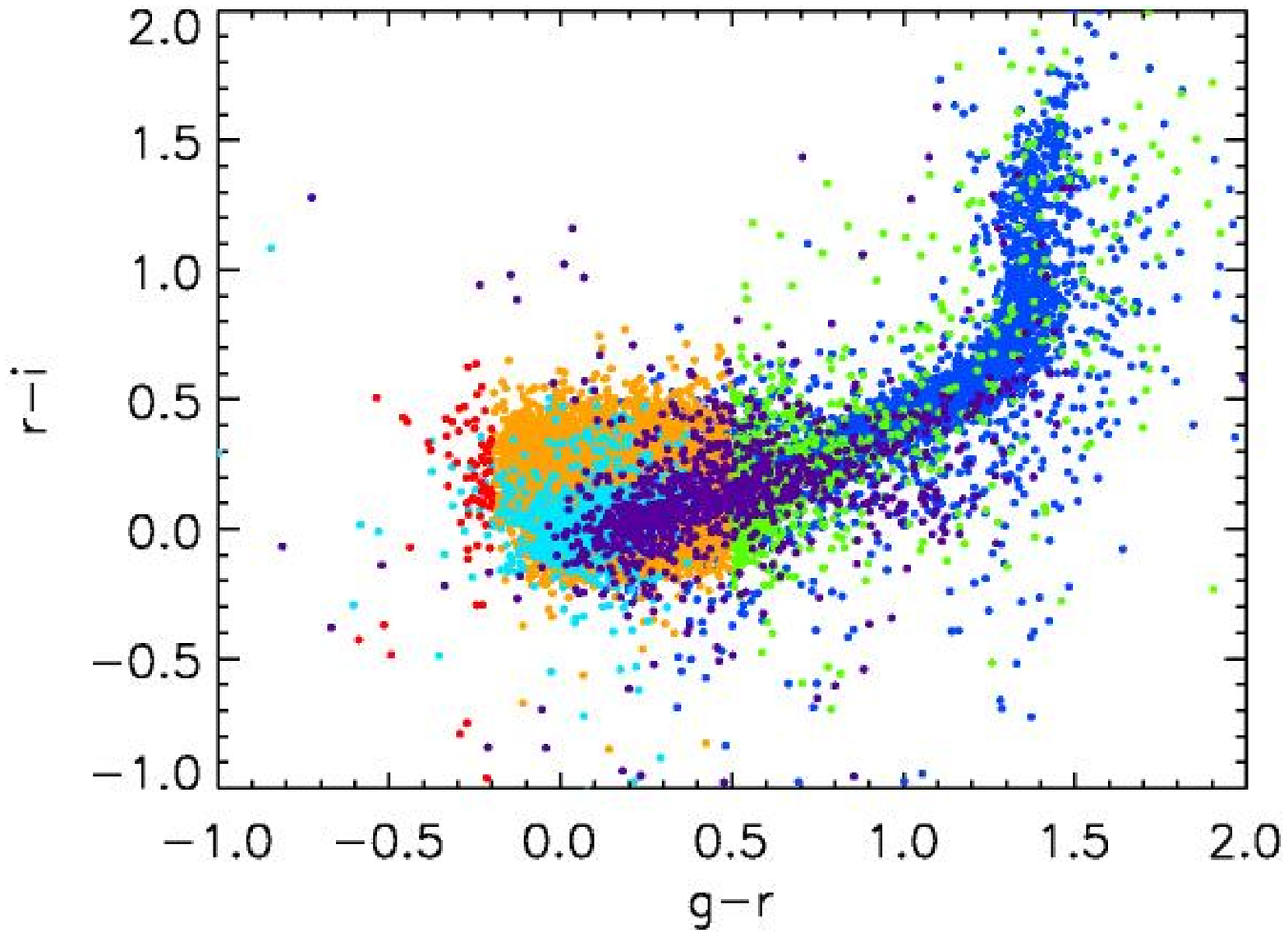,angle=0,width=0.4\linewidth}
\caption{Colour-colour plots for 4\,648 variables brighter than $g =
  19.0$ (left) and 21\,789 variables brighter than $g = 22.0$ (right). The
  upper panels are $g-r$ versus $u-g$, the lower panels are $r-i$
  versus $g-r$. Sesar et al. (2007) label these regions as white
  dwarfs (Region 1, red), low-redshift quasars (Region 2, orange), M
  dwarf/white dwarf binaries (Region 3, green), RR Lyraes (Region
  4, cyan), main stellar locus (Region 5, blue) and high-redshift
  quasars (Region 6, purple).}
\label{fig:colcol}
\end{figure*}

\begin{table*}
\centering
\begin{tabular}{llrrrrrrrrr}
\hline
\multicolumn{2}{c}{\null} & \multicolumn{2}{c}{$g < 19$} & \multicolumn{2}{c}{$g < 20.5$} & \multicolumn{2}{c}{$g < 22$} \\
\hline
Region & Name & \% all & \% var & \% all & \% var & \% all & \% var  \\
1 (red)  & white dwarfs          &  0.09 &  0.43 &  0.16 &  0.33 &  0.20 &  0.36 \\
2 (orange)  & low-redshift quasars  &  0.29 & 21.32 &  1.20 & 52.96 &  5.31 & 61.05 \\
3 (green)  & dM/WD pairs           &  5.95 &  1.01 &  9.51 &  1.59 & 11.77 &  3.65 \\
4 (cyan)  & RR Lyrae stars        &  3.27 & 15.23 &  3.62 &  9.31 &  3.47 &  7.85 \\
5 (blue)  & stellar locus stars   & 75.93 & 56.41 & 74.62 & 31.44 & 69.65 & 22.48 \\
6 (purple)  & high-redshift quasars & 14.47 &  5.59 & 10.89 &  4.37 &  9.61 &  4.62 \\
\hline
    & total count & 283,899 & 4,648 & 447,800 & 12,788 & 518,058 & 21,789 &
\end{tabular}
\caption{The distribution of candidate variable sources in the $g-r$ 
  versus $u-g$ diagram. The columns list the fraction of the whole sample 
  and the variable subsample lying in the six regions of the
  colour-colour plot.\label{tab:stats}}
\end{table*}

\citet{Iv07} constructed a catalogue of one-million standard stars
with $r$ magnitudes 14.0-22.0, by averaging repeated
observations of unresolved sources in the 290\,deg$^2$ area of
Stripe 82.  \citet{Se07} used the catalogue to carry out the
first analysis of variability in Stripe 82. In particular, they
applied cuts $\chi^2_r > 3$ and $\chi^2_g > 3$, followed by the
requirement that the root-mean-square scatter exceeded 0.05\,mag, to
identify variability, obtaining a catalogue of 20\,533 variable
sources.

Bramich et al.'s (2008) light-motion curve catalogue (LMCC) is based
on observations of Stripe 82 restricted to a smaller area of
249\,deg$^2$, extending in right ascension from $20.7^{\mbox{\small
h}} < \alpha < 3.3^{\mbox{\small h}}$ with a width 2\fdg52 in
declination from $\delta = -1\fdg26$ to $1\fdg26$. We extracted a
sample of high-quality variable stars from the LMCC by imposing the
restrictions that: i) $\chi^2_r >3$ and $\chi^2_g >3$, ii) a cut on
the Stetson index $L_g >1$, iii) at least 10 good epochs are retained,
giving a a catalogue of 21\,939 variable objects. Applying Sesar et
al's cuts to our catalogue gives 22\,483 objects, with $\approx 80 \%$
in common with our subsample based on Stetson index cuts. Even though
Sesar et al.'s cuts give more candidates, the additional objects
possess variability in different passbands that is not
well-correlated.

\begin{figure*}
\epsfig{file=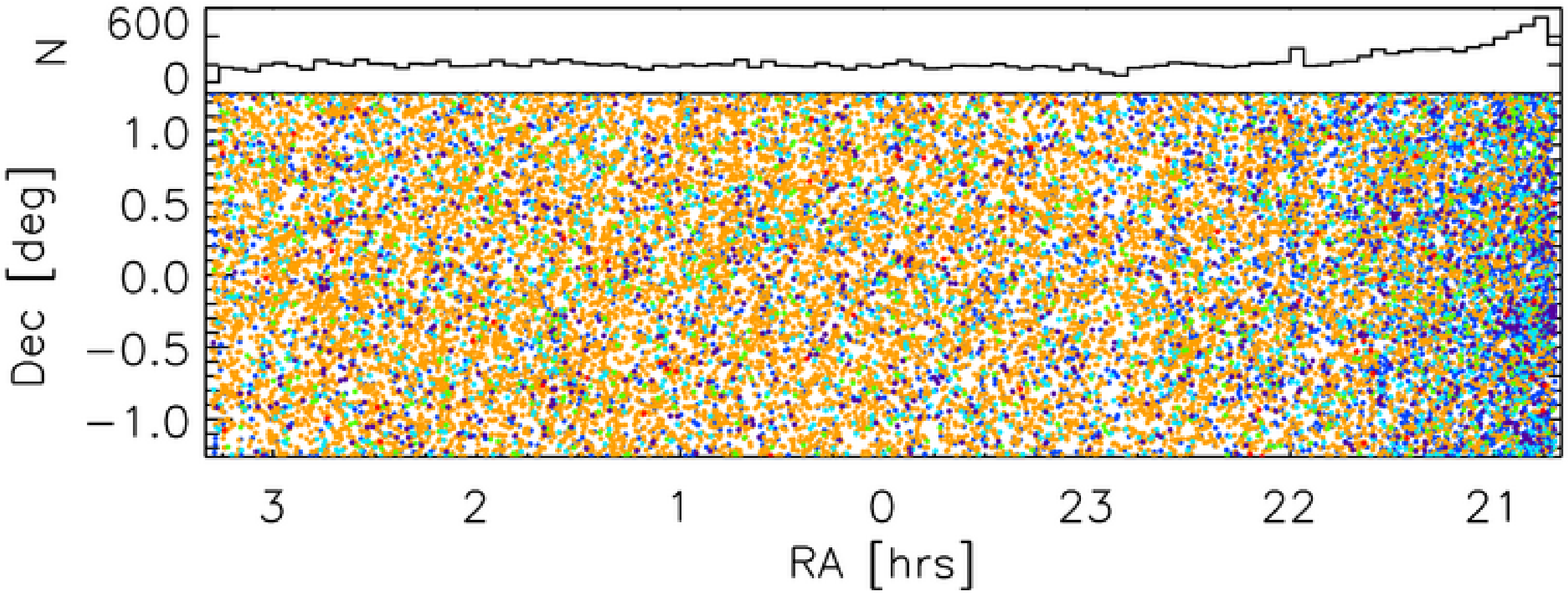,angle=0,width=\linewidth}
\caption{The spatial distribution of the variable subsample in Stripe
  82. Objects are colour-coded according to the Regions of the
  colour-colour plot in which they lie (see
  Figure~\ref{fig:colcol}). The upper panel shows the number of all
  the variable objects versus right ascension.}
\label{fig:spatial}
\end{figure*}

Sesar and co-workers used a colour-colour plot to discriminate between
different classes of variable objects.  In Figure~\ref{fig:colcol},
our stellar subsample is plotted in $g-r$ versus $u-g$ and $r-i$
versus $g-r$. Here, following \citet{Se07}, the $g-r$ versus $u-g$
plot is divided into six regions, and labelled according to possible
occupants: white dwarfs (the red-coloured Region 1), low-redshift
quasars (the orange-coloured Region 2), M dwarf/white dwarf binaries
(the green-coloured Region 3), RR Lyraes (the cyan-coloured Region 4),
stellar locus stars (the blue-coloured Region 5) and high-redshift
quasars (the purple-coloured Region 6). The colour-space divisions
provide only very rough classifications. In some cases (such as region
1), the label does not even describe the typical population, and we
merely use the labels as a point of comparison to Sesar's work.

The percentages of the variable subsample and the whole sample lying
in the regions of the colour-colour plots are given in
Table~\ref{tab:stats}, whilst sample lightcurves have already been
shown in Figure~\ref{fig:lcs}.  Almost all ($>$93\,per cent) of the variable
objects lie in three regions -- namely, low-redshift quasars (53 per cent
of the catalogue), stellar locus stars (31.4\,per cent), and RR Lyrae
stars (9.3\,per cent). When split according to magnitude, the bright ($g$$<$
19.0) variable-sky is dominated by stellar locus stars, but the faint
($g$$<$22.0) variable-sky is dominated by quasars. We can compare our
results to Table 1 of \citep{Se07}, which shows the same quantities
for their variable subsample. Our variability criteria picks out more
variable objects, and in particular more denizens of the main stellar
locus.

The spatial distribution of variable objects in Stripe 82 is shown in
Figure~\ref{fig:spatial}. The equatorial stripe reaches down to low
Galactic latitudes beyond $\alpha \approx 18^{\rm h}$ (see e.g.
Figure 1 of Belokurov et al. 2007). Variables belonging to Region 4
(RR Lyraes) and Region 5 (the main stellar locus) dominate here,
whereas variables belonging to the other Regions are more uniformly
dispersed in right ascension.
\label{lastpage}
\end{document}